\documentclass[11pt]{article}

\PassOptionsToPackage{nosort}{cite}

\usepackage{graphicx}
\usepackage{float}
\usepackage{mathtools}
\usepackage{scalerel}
\usepackage{amssymb}
\usepackage{amsfonts}
\usepackage{dsfont}
\usepackage{mathrsfs}
\usepackage{enumitem}
\usepackage{multirow}
\usepackage[font=small,format=hang,labelfont={sf,bf}]{caption}
\usepackage{chet}
\usepackage{subcaption}
\newcommand{\lsp}{\hspace{1pt}}

\newcommand{\lnsp}{\hspace{-1pt}}

\renewcommand{\geq}{\geqslant}

\newcommand{\MN}{M\hspace{-0.25pt}N\xspace}
\newcommand{\MNmath}{\ensuremath{M\lnsp N}}
\newcommand{\TotS}{T\hspace{-1pt}o\hspace{0.3pt}tS}

\allowdisplaybreaks

\definecolor{darkblue}{rgb}{0.1,0.1,0.7}
\hypersetup{colorlinks,
           linkcolor={darkblue},
           citecolor={darkblue},
           urlcolor={darkblue}
}

\date{May 2022}

\title{Bootstrapping Mixed MN Correlators in 3D}

\author{Stefanos R.\ Kousvos$^{a,b,c}$ and Andreas Stergiou$^{d,e}$}

\affiliation{
$^a$Department of Physics, University of Crete, Heraklion
GR-70013, Greece\\
$^b$Institute of Theoretical and Computational Physics (ITCP),
Department of Physics,\\\vspace{-3pt}
University of Crete, 70013 Heraklion, Greece\\
$^c$ Department of Physics, University of Pisa and INFN,
Largo Pontecorvo 3, I-56127 Pisa, Italy\\
$^d$Theoretical Division, MS B285, Los Alamos National Laboratory, Los
Alamos, NM 87545, USA\\
$^e$Department of Mathematics, King's College London,\\\vspace{-3pt}
Strand, London WC2R 2LS, United Kingdom}

\abstract{The recent emergence of the modern conformal bootstrap method for
the study of conformal field theories (CFTs) has enabled the revisiting of
old problems in classical critical phenomena described by three-dimensional
CFTs. The study of such CFTs with $O(m)^n \rtimes S_n$ global symmetry,
also known as \MN models, is pursued in this work. Systems of mixed
correlators involving scalar operators in two different representations of
the global symmetry group are considered.  Isolated allowed regions are
found in parameter space for various values of $m$ and $n$.  These
``islands'' can be separated into two qualitative groups: those close to
the unitarity bound and those further away. As a by-product of our analysis
generic tensor structures required to bootstrap any $G^n \rtimes S_n$
theory with $G$ arbitrary are worked out.}

\begin{document}

\maketitle

\toc

\newsec{Introduction}
Since its modern revival, the numerical conformal bootstrap
\cite{Rattazzi:2008pe} has provided a powerful and efficient tool for the
study of conformal field theories (CFTs).\foot{For a review see
\cite{Poland:2018epd} and for a pedagogic introduction see
\cite{Chester:2019wfx}.} Its most remarkable successes have been found in
$d=3$ spacetime dimensions, where it has been used to determine critical
exponents and other quantities of interest in the $O(N)$ family of models,
\cite{Kos:2016ysd} ($N=1$), \cite{Chester:2019ifh} ($N=2$) and
\cite{Chester:2020iyt} ($N=3$), as well as the supersymmetric Ising ($N=1$)
variant \cite{Rong:2018okz, Atanasov:2018kqw}.

Beyond these widely studied examples, it is desirable to extend
applications to less examined ones.  Three-dimensional multi-scalar
theories provide ideal candidates, since they are simultaneously some of
the simplest non-trivial theories one may write down, but at the same time
they are also physically relevant due to their applications to classical
critical phenomena. If one wishes to scan over all the low lying scalar
operators as externals in the theory (which would presumably make results
more constraining), the number of sum rules one needs to consider is (much)
larger than the $O(N)$ case.  This is due to the fact that subgroups of
$O(N)$ always have (the same or) more irreducible representations than
$O(N)$ in the relevant operator product expansions (OPEs).\foot{Recently
there has been some notable progress regarding techniques for higher
dimensional scans in parameter space; see e.g.\ \cite{Reehorst:2021ykw}.
See also \cite{Reehorst:2021hmp} for an example with two externals but many
exchanged parameters scanned over.} Additionally, one of the most important
assumptions in the $O(N)$ case, namely $\Delta_{\phi^\prime} \geq 3$ with
$\phi^\prime$ the next-to-leading operator in the vector representation of
$O(N)$, cannot be used since in a typical multi-scalar theory the
next-to-leading operator in the defining representation has a strongly
relevant dimension. This is because the vector representation ``$\phi^3$''
operator in a generic multi-scalar theory cannot be neglected due to the
equation of motion as in the $O(N)$ case, which in turn limits the size of
the gaps we can impose in this sector.

There are many three-dimensional multi-scalar models to consider. Perhaps
the most well-known one is the cubic model, which is a theory whose global
symmetry group consists of the cubic subgroup $\mathbb{Z}_2^{\;\,3}\rtimes
S_3$ of $O(3)$. Conformal field theories with cubic and the more general
hypercubic symmetry have been bootstrapped in~\cite{Rong:2017cow,
Stergiou:2018gjj, Kousvos:2018rhl, Kousvos:2019hgc}.  Many other examples
to be considered include fixed points already discovered with the
$\varepsilon$ expansion a long time ago,\foot{See for example
\cite{Rychkov:2018vya} for a relatively recent review of known fixed points
in the $\varepsilon$ expansion below four dimensions.} as well as a wealth
of new $\varepsilon$ expansion fixed points presented
in~\cite{Osborn:2020cnf}. With the addition of fermions there are
presumably many fixed points to study with the bootstrap, including
supersymmetric ones~\cite{Liendo:2021wpo}.

In this work we continue the study of scalar \MN models by considering a
multi-correlator study of CFTs with $O(m)^n \rtimes S_n$ global symmetry in
three dimensions.  These models have been analyzed with the $\varepsilon$
expansion~\cite{Mukamel, Shpot, Shpot2, Mudrov, Mudrov2, Osborn:2017ucf,
Rychkov:2018vya} as well as the numerical conformal
bootstrap~\cite{Stergiou:2019dcv}, where a single-correlator study was
pursued. Our goal is to find the minimal set of conditions that allow
us to isolate allowed regions in parameter space at the positions of kinks
of the single-correlator bounds obtained in~\cite{Stergiou:2019dcv}. As was
explored in~\cite{Stergiou:2019dcv} and also in~\cite{Henriksson:2021lwn},
bounds of $m=2$ theories contain two kinks with potential applications to
critical phenomena for low $n$ and of theoretical interest at larger $n$.
There is no proof that a kink of the type we find in a numerical bootstrap
bound must be due to the presence of an actual CFT. However, we will
consider such kinks to be strong indicators for the presence of actual
CFTs. For an in-depth discussion of experimental realizations of the CFTs we will study and their significance within field theory we refer the reader to \cite{Stergiou:2019dcv}.

By considering systems of correlators involving operators in multiple
irreducible representations (irreps) of the global symmetry group, we
manage to find isolated allowed regions (islands) at the positions of kinks
of single-correlator bounds. To get these islands we make assumptions on
the operator spectrum, but we try to keep the number of such assumptions to
a minimum. We find that we typically have to require the presence of a conserved current
and stress-energy tensor and gaps to the next-to-leading spin-1 and spin-2
operators in the corresponding irreps. Also, if we wish to obtain an island
in the $\Delta_\phi$-$\Delta_S$ plane,\footnote{Where $\phi$ is the order
parameter field and $S$ the scalar singlet.} we need to require that the
leading operator in a specific irrep (called $X$ below) saturates the bound
obtained from the single-correlator bootstrap. However, to obtain an island
in the $\Delta_\phi$-$\Delta_X$ plane, no such assumption is required. We note that the $\Delta_X$ bound is of interest since it displays pronounced features/kinks in parameter space.

One particularly interesting island is found for the second kink of the
$\MNmath_{20,2}$ theory (see Fig.~\ref{MN_20_2_island} below). This island
is obtained in the $\Delta_\phi\text{-}\Delta_S$ plane, where $\phi$ is the
defining and $S$ the singlet irrep of $\MNmath_{20,2}$. A notable feature
of this island is that the dimensions $\Delta_\phi$ and $\Delta_S$ that
define it are low, while the assumptions we make to obtain it force the
dimensions of next-to-leading operators in a variety of irreps to be high.

For the experimentally relevant $\MNmath_{2,2}$ and $\MNmath_{2,3}$
theories we find islands at the first kinks of the corresponding
single-correlator bootstrap bounds. One would expect the values of
$\Delta_\phi$ and $\Delta_S$ in these islands to match the results obtained
in the $\varepsilon$ expansion. This is not the case, however, and the
potential resolutions are either that the $\varepsilon$ expansion results
in the $\varepsilon\to1$ limit are not trustworthy or that the bootstrap
kink is due to a theory distinct from that of the $\varepsilon$ expansion.
We note that at large $m$ and large $n$ the bootstrap results we obtain for
$\MNmath_{m,n}$ theories agree very well with the $\varepsilon$ expansion.

For various values of $m$ and $n$ we initiate a preliminary study of
symmetric tensors $Z^{ab}_{ij}$ of $\MNmath_{m,n}$, similar in spirit to
\cite{Reehorst:2020phk}. We find a family of kinks that in the large $n$
limit converge to the point $\Delta_Z = \frac12$ and $\Delta_Y = 2$, which
hints towards a standard Hubbard--Stratonovich description in this limit.
Conversely, in the large $m$ limit, the field $Y$ does not converge to
$\Delta_Y = 2$. We delegate a careful study of these bounds to future work.

We also work out the required tensor structures (under global symmetry) for
generic groups $G^n \rtimes S_n$ with $G$ arbitrary. There are at least two
motivations for doing so. First, there exist experimentally relevant cases
among such models. For example, beyond the cubic ($\mathbb{Z}_2^{\;\,3}
\rtimes S_3$), $MN_{2,2}$ ($O(2)^2\rtimes S_2$) and $MN_{2,3}$
($O(2)^3\rtimes S_3$) models mentioned earlier, there are also the so
called tetragonal theories ($D_4{\!}^n\rtimes S_n$, with $D_4$ the
dihedral group of 8 elements). An incomplete list of references where the
reader may find a number of physically motivated examples is
\cite{Pelissetto:2000ek,PhysRevLett.34.481,PhysRevB.13.5065,PhysRevB.13.5078,PhysRevB.13.5086}.
A second, broader, motivation stems from our pursuit to better understand
the space of three-dimensional CFTs. We hope that our results will enable a
variety of future bootstrap studies.

This paper is organized as follows. In the next section we present a quick
review of \MN models as understood from a Lagrangian point of view and
describe our results for global symmetries of the type $G^n\rtimes S_n$. In
section \ref{numerics} we present our numerical bootstrap results. We
conclude in section \ref{conc}. Four technical appendices include
calculations of crossing equations in a variety of cases relevant for this
work.

\newsec{\MN and \texorpdfstring{$\boldsymbol{G^n \rtimes
S_n}$}{G\^{}n rtimes\xspace S\_n} symmetries}[rev]
\subsec{Definition and Lagrangian description in
\texorpdfstring{$d=4-\varepsilon$}{d=4-epsilon}}
The global symmetry group $\MNmath_{m,n}=O(m)^n \rtimes S_n$ consists of
$n$ copies of the $O(m)$ model that can be permuted with each other.
Consider a scalar field multiplet $\phi^a_i$, where the upper index labels
the copy and the lower index is an $O(m)$ index.  With this field we may
write the multi-scalar $\MNmath_{{m,n}}$-invariant Lagrangian
\begin{equation}
\mathscr{L}=\partial_\mu \phi^a_i\lsp \partial^\mu \phi^a_i
+ m^2 \lsp \phi^a_i \phi^a_i
+ u\lsp (\phi^a_i \phi^a_i)^2
+ v\lsp \delta_{abcd}\lsp\phi^a_i \phi^b_i \phi^c_j \phi^d_j\,,
\label{eq1}
\end{equation}
where summation over repeated indices is implicit and we keep up to quartic
terms in $\phi$. The field $\phi^a_i$ transforms in the ``defining''
representation of $\MNmath_{m,n}$. In other words, the index $a$ has its
values permuted by $S_n$, and $O(m)$ acts as $\phi^a_i \rightarrow
R_{ij}\phi^a_j$ with $R$ including proper and improper rotations. If one
were to set $u=0$ in \eqref{eq1}, then \eqref{eq1} would be the Lagrangian
of $n$ decoupled $O(m)$ models. Thus, when both $u$ and $v$ are non-zero we
obtain a theory describing $n$ coupled $O(m)$ models. Single-coupling
deformations of the $O(N)$ theory lead to a total of four fixed
points~\cite{Brezin:1973jt, Osborn:2017ucf, Rychkov:2018vya}, which
correspond to the free theory, $n$ coupled $O(m)$ models (i.e.\ the \MN
theory), $n$ decoupled $O(m)$ models, and the $O(mn)$ theory.

The generalization of \eqref{eq1} to the $G^n \rtimes S_n$-symmetric case
for $G$ arbitrary is straightforward:
\begin{equation}
\mathscr{L}=\partial_\mu \phi^a_i\lsp \partial^\mu \phi^a_i
+ m^2 \lsp\phi^a_i \phi^a_i
+ u\lsp (\phi^a_i \phi^a_i)^2
+ \sum_r v_r \lsp\delta_{abcd}\lsp T^r_{ijkl}\lsp
\phi^a_i \phi^b_j \phi^c_k \phi^d_l\,,
\label{eq2}
\end{equation}
where $r$ labels the four-index fully symmetric invariant tensors
$T^r_{ijkl}$ of $G$, to each of which we associate a coupling $v_r$. Only
fields from the same copy of $G$ interact when $u=0$. In tetragonal
theories (see e.g.\ \cite{Rychkov:2018vya}), where $G=D_4$ with $D_4$ the
dihedral group of eight elements, we have $r=1,2$.

A minor comment is that instead of viewing \eqref{eq1} and \eqref{eq2} as
$n$ decoupled $G$-symmetric theories which are consequently coupled via the
addition of the $u$ term, we may also view them as deformations of an
$O(mn)$ symmetric theory with a term that breaks the symmetry down to $G^n
\rtimes S_n$. Both approaches are fruitful. For example, the authors of
\cite{Komargodski:2016auf} study hypercubic theories, where
$G=\mathbb{Z}_2$, by performing conformal perturbation theory around the
CFT of $n$ decoupled Ising models, whereas the authors of
\cite{Chester:2020iyt} discuss the cubic theory, where $G=\mathbb{Z}_2$ and
$n=3$, by writing it as a deformation of $O(3)$ and then using conformal
perturbation theory. Both approaches led to different and interesting
results.

\subsec{The \texorpdfstring{$\phi^a_i\times
\phi^b_j$}{phi\^{}a\_i\xspace\texttimes\xspace phi\^{}b\_j} OPE}
The product of two fields transforming in the defining representation of
$G^n \rtimes S_n$ is schematically expressed as
\begin{equation}
\phi^a_i \times \phi^b_j \sim \delta^{ab}\delta_{ij}S
+\delta_{ij}X^{ab}+{I_1}^{ab}_{ij} + \cdots + {I_\mathcal{N}}^{ab}_{ij} +
Z^{ab}_{ij} + B^{ab}_{ij}\,.
\label{fullMNope}
\end{equation}
Here the lower indices take on values $i,j=1,\ldots,m$ with $m$ the
dimension of the defining irrep of $G$ and the upper indices take on values
$a,b=1,\ldots,n$.\foot{For every value $a_*$ of the upper index $a$,
$\phi_i^{a_*}$ furnishes the defining irrep of the corresponding $G$ in
$G^n$.} The representations on the right-hand side of \eqref{fullMNope} are
to be understood as follows.  The last two representations do not exist if
$a=b$, whereas all other representations exist only if $a=b$. The first
representation, $S$, is the singlet. The representation $X$ is a singlet of
$G^n$ and traceless in $S_n$, or to rephrase, it is the fundamental
representation of $S_n$.  The representations $Z$ and $B$ satisfy
$Z^{ab}_{ij}= Z^{ba}_{ji}$ and $B^{ab}_{ij}=-B^{ba}_{ji}$. The remaining
irreps $I_i, i=1,\ldots,\mathcal{N}$, correspond to non-singlet
representations of $G$.  The dimensions of the irreps $(S, X, I_1, \ldots,
I_\mathcal{N}, Z, B)$ are $(1, n-1, n \lsp \dim R_1, \ldots, n\lsp \dim
R_{\mathcal{N}}, m^2\frac{n(n-1)}{2}, m^2\frac{n(n-1)}{2})$, where
$R_1,\ldots, R_{\mathcal{N}}$ are the non-singlet irreps of $G$ that appear
in the $\phi_i\times\phi_j$ OPE of $G$.  An easy mnemonic rule for
obtaining the decomposition in \eqref{fullMNope} is to decompose onto
irreps of $G$ when $a=b$ and to simply symmetrize and antisymmetrize the
indices when $a \neq b$.

We may decompose a four point function $\langle \phi^a_i \phi^b_j \phi^c_k
\phi^d_l \rangle$ into the irreps that appear in \eqref{fullMNope}. Doing
this, we find tensor structures of the global symmetry which in turn
determine the sum rules to be eventually studied numerically. Notably,
these tensor structures are projectors.\footnote{Assuming they are normalized properly, which we won't always do. Nevertheless, by slight abuse of terminology we will still call them projectors.} The explicit expressions are
\begin{align}
(P^S)^{abcd}_{ijkl}&=
\frac{1}{mn}\delta^{ab}\delta^{cd}\delta_{ij}\delta_{kl}\,,\nonumber\\
(P^X)^{abcd}_{ijkl}&=
  \frac{1}{m}\delta_{ij}\delta_{kl}\Big(
  \delta^{abcd}-\frac{1}{n}\delta^{ab}\delta^{cd}\Big)\,,\nonumber\\
(P^{I_1})^{abcd}_{ijkl}&=
  \delta^{abcd}\lsp{P^{R_1}}_{\!ijkl}\,,\nonumber\\
&\ldots\nonumber\\
(P^{I_\mathcal{N}})^{abcd}_{ijkl}&=
  \delta^{abcd}\lsp{P^{R_\mathcal{N}}}_{\!ijkl}\,,\nonumber\\
(P^Z)^{abcd}_{ijkl}&=
  \tfrac{1}{2} ( ( \delta^{ac}\delta^{bd}-\delta^{abcd})
  \delta_{ik}\delta_{jl} + ( \delta^{ad}\delta^{bc} -
 \delta^{abcd})\delta_{il}\delta_{jk} )\,, \nonumber\\
(P^B)^{abcd}_{ijkl}&=
  \tfrac{1}{2} ( ( \delta^{ac}\delta^{bd} -\delta^{abcd})
  \delta_{ik}\delta_{jl} - ( \delta^{ad}\delta^{bc} -
  \delta^{abcd})\delta_{il}\delta_{jk} )\,,
\label{generalprojectors}
\end{align}
where the tensors $P^{R_1}{}_{\!ijkl}, \ldots,
P^{R_{\mathcal{N}}}{}_{\!ijkl}$ correspond to the projectors of the irreps
$R_1, \ldots, R_{\mathcal{N}}$ of $G$. Note that all invariant tensors are
expressed in terms of (generalized) Kronecker deltas, up to the form of the
$P^{R_1}{}_{\!ijkl}, \ldots, P^{R_{\mathcal{N}}}{}_{\!ijkl}$ invariant
tensors.

We now give a set of explicit examples to clear up any confusion.

\subsubsec{Example 1: hypercubic theories}
We start with hypercubic theories, which have been bootstrapped in
\cite{Rong:2017cow, Stergiou:2018gjj, Kousvos:2018rhl, Kousvos:2019hgc}. In
this example, $G=\mathbb{Z}_2$ and so the lower indices ($i,j,k,l$) may be
dropped.  Also, in the OPE between two operators charged under
$\mathbb{Z}_2$ only the singlet representation appears, thus we have no
``$I$'' representations. Hence, the projectors are
\begin{align}
\begin{split}
(P^{S})^{abcd}&=\frac{1}{n}\delta^{ab}\delta^{cd}\,,\\
(P^X)^{abcd}&=\delta^{abcd}-\frac{1}{n}\delta^{ab}\delta^{cd}\,, \\
(P^Z)^{abcd}&=
  \tfrac{1}{2} ( ( \delta^{ac}\delta^{bd} -\delta^{abcd}) +
    ( \delta^{ad}\delta^{bc} - \delta^{abcd}) )\,, \\
(P^B)^{abcd}&=
  \tfrac{1}{2} ( ( \delta^{ac}\delta^{bd} -\delta^{abcd}) -
  (\delta^{ad}\delta^{bc} - \delta^{abcd}) )\,,
\end{split}
\label{generalprojectors1}
\end{align}
where the irreps $(S,X,Z,B)$ have dimensions
$(1,n-1,\frac{n(n-1)}{2},\frac{n(n-1)}{2})$, in agreement with the
dimensions stated above when $m=1$.

\subsubsec{Example 2: \MN theories}
Next, we consider the \MN theories studied in \cite{Stergiou:2019dcv} and
\cite{Henriksson:2021lwn}. In this example we have $G=O(m)$. Since the OPE
between two vectors of $O(m)$ exchanges, beyond the singlet, the
two-index traceless symmetric $T$ and antisymmetric $A$ irreps, we have
$R_1=T$ and $R_2=A$. For these irreps we know that
$P^{R_1}{}_{\!ijkl}=\frac{1}{2}(\delta_{ik}\delta_{jl}+\delta_{il}\delta_{jk}-\frac{2}{m}\delta_{ij}\delta_{kl})$
and $P^{R_2}{}_{\!ijkl}=\frac{1}{2}(\delta_{ik}\delta_{jl}-\delta_{il}\delta_{jk})$. We thus obtain
\begin{align}
\begin{split}
(P^S)^{abcd}_{ijkl}&=
  \frac{1}{mn}\delta^{ab}\delta^{cd}\delta_{ij}\delta_{kl}\,,\\
(P^X)^{abcd}_{ijkl}&=\frac{1}{m}\delta_{ij}\delta_{kl}\Big(
  \delta^{abcd}-\frac{1}{n}\delta^{ab}\delta^{cd}\Big)\,, \\
(P^{I_1})^{abcd}_{ijkl}&=
  \delta^{abcd}\lsp{P^{R_1}}_{\!ijkl}\,,\\
(P^{I_2})^{abcd}_{ijkl}&=
  \delta^{abcd}\lsp{P^{R_2}}_{\!ijkl}\,,\\
(P^Z)^{abcd}_{ijkl}&=
  \tfrac{1}{2} ( ( \delta^{ac}\delta^{bd} -\delta^{abcd})
  \delta_{ik}\delta_{jl} + ( \delta^{ad}\delta^{bc} -
  \delta^{abcd})\delta_{il}\delta_{jk} )\,, \\
(P^B)^{abcd}_{ijkl}&=
  \tfrac{1}{2} ( ( \delta^{ac}\delta^{bd} -\delta^{abcd})
  \delta_{ik}\delta_{jl} - ( \delta^{ad}\delta^{bc} -
  \delta^{abcd})\delta_{il}\delta_{jk} )\,,
\end{split}
\label{generalprojectors2}
\end{align}
where the representations $(S, X, I_1, I_2, Z, B)$ have dimensions
\eqn{(1, n-1, n \frac{(m-1)(m+2)}{2}, n\frac{m(m-1)}{2},
m^2\frac{n(n-1)}{2}, m^2\frac{n(n-1)}{2})\,,}[]
in agreement with the general formulas given earlier and the computations
of \cite{Stergiou:2019dcv}. In what follows we will rename $I_1$ and $I_2$
as $Y$ and $A$ to keep the same names as \cite{Stergiou:2019dcv}.

\subsubsec{Example 3: Tetragonal theories and their generalizations}
We may also study a generalization of the tetragonal theories bootstrapped
in \cite{Stergiou:2019dcv}. In this example we have $G=\mathbb{Z}_2^{\;\,m}
\rtimes S_m$. Thus, similarly,
\begin{align}
\begin{split}
(P^S)^{abcd}_{ijkl}&=
  \frac{1}{mn}\delta^{ab}\delta^{cd}\delta_{ij}\delta_{kl}\,,\\
(P^X)^{abcd}_{ijkl}&=
  \frac{1}{m}\delta_{ij}\delta_{kl}\Big(\delta^{abcd}
  -\frac{1}{n}\delta^{ab}\delta^{cd}\Big)\,, \\
(P^{I_1})^{abcd}_{ijkl}&=
  \delta^{abcd}\Big(\delta_{ijkl}-\frac{1}{m}\delta_{ij}\delta_{kl}\Big)\,,\\
(P^{I_2})^{abcd}_{ijkl}&=
  \delta^{abcd}\lsp\tfrac{1}{2}(\delta_{ik}\delta_{jl}
  +\delta_{il}\delta_{jk}-2\delta_{ijkl})\,,\\
(P^{I_3})^{abcd}_{ijkl}&=
  \delta^{abcd}\lsp \tfrac{1}{2}(\delta_{ik}\delta_{jl}-
  \delta_{il}\delta_{jk})\,,\\
(P^Z)^{abcd}_{ijkl}&=
  \tfrac{1}{2} ( ( \delta^{ac}\delta^{bd} -\delta^{abcd})
  \delta_{ik}\delta_{jl} + ( \delta^{ad}\delta^{bc} -
  \delta^{abcd})\delta_{il}\delta_{jk} )\,,\\
(P^B)^{abcd}_{ijkl}&=
  \tfrac{1}{2} ( ( \delta^{ac}\delta^{bd} -\delta^{abcd})
  \delta_{ik}\delta_{jl} - ( \delta^{ad}\delta^{bc} -
  \delta^{abcd})\delta_{il}\delta_{jk} )\,,
\end{split}
\label{generalprojectors3}
\end{align}
where now the irreps $(S, X, I_1, I_2, I_3, Z, B)$ have dimensions
\eqn{(1, n-1, n (m-1), n\frac{m(m-1)}{2}, n\frac{m(m-1)}{2},
m^2\frac{n(n-1)}{2}, m^2\frac{n(n-1)}{2})\,.}[]
These indeed agree with and generalize to arbitrary $m$ the results of
\cite{Stergiou:2019dcv}.

\subsec{The \texorpdfstring{$\phi^a_i\times
X^{bc}$}{phi\^{}a\_i\xspace\texttimes\xspace X\^{}bc} OPE}
In our mixed correlator system we will need to consider the OPE of $\phi$
with $X$. A first observation is that the decomposition of this OPE does
not depend on $G$ because $X$ is a singlet of $G$. We also know that $\phi
\times X$ should exchange $\phi$, as dictated by self-consistency of the
three-point function, i.e.\ calculating the three point function with
different OPEs should give the same result. Lastly, the existence of the
defining representation ($\phi$) on the right hand side of the OPE implies
the existence also of an antisymmetric representation ($\bar{A}$) on the
right-hand side, see e.g.\ \cite[Chapter 4.3]{Kousvos:2021mpw}. If one now counts the
dimensions of these two irreps ($\phi$ and $\bar{A}$), their sum turns out
equal to the product of the dimensions on the left-hand side of the OPE.
Thus, the OPE is fully decomposed. Schematically, the decomposition looks
like
\begin{equation}
\phi^a_i X^{bc} \sim (P^\phi)^{abcdef} \phi^d_i X^{ef} +
(P^{\bar{A}})^{abcdef} \phi^d_i X^{ef}\,,
\label{aeq1}
\end{equation}
where
\begin{equation}
(P^\phi)^{abcdef}=\delta_{abcdef}-\frac{1}{n}(\delta^{bc}\delta^{adef}+\delta^{ef}\delta^{abcd})+\frac{1}{n^2}\delta^{ad}\delta^{bc}\delta^{ef}
\label{aeq7}
\end{equation}
and
\begin{equation}
(P^{\bar{A}})^{abcdef}=-\delta^{abcdef}+\frac{n-1}{n}\delta^{ad}\delta^{bcef}
+\frac{1}{n}(\delta^{bc}\delta^{adef}+\delta^{ef}\delta^{abcd})-\frac{1}{n}\delta^{ad}\delta^{bc}\delta^{ef}\,.
\label{aeq6}
\end{equation}
It is a straightforward exercise to contract the tensors quoted above in
\eqref{aeq1} and obtain the expected representation dimensions. Note that
\eqref{aeq7} and \eqref{aeq6} are the projectors we need in order to obtain
the sum rules from the corresponding crossing equations. An additional
check one can perform is test the above decomposition for some discrete
group $G$ in which the decompositions are known, and then remembering that
the formula is $G$-independent one has proved the result.

\subsec{The \texorpdfstring{$X^{ab}\times
X^{cd}$}{X\^{}ab\xspace\texttimes\xspace X\^{}cd} OPE}
It is useful to remember at this point that the $X$ representation is
simply the fundamental representation of $S_n$, and so its OPE
decomposition is already known from e.g.\ \cite{Hogervorst:2016itc}; see
also the later works \cite{Rong:2017cow} and \cite{Stergiou:2018gjj}.
Hence, both projectors and crossing equation sum rules are known. Note that
the fundamental of $S_n$ is usually written with one index. For the
explicit map between the one- and two- index notations see
\cite{Kousvos:2021mpw}.

\subsec{The \texorpdfstring{$n\geq 4$}{n>= 4}
\texorpdfstring{$Z^{ab}_{ij} \times Z^{cd}_{kl}$}{Z\^{}ab\_ij\xspace
\texttimes\xspace Z\^{}cd\_kl} OPE}
In this work we will also consider the $\langle ZZZZ\rangle$ single correlator bootstrap
for values of $n\geq 4$ (and $m\geq 2$). This is because we can then probe various large
parameter limits, such as the $m \rightarrow \infty$ and $n \rightarrow
\infty$ limits.\footnote{For $m \geq 2$ and $n \geq 4$ the sum
rules take their most general form, valid for any such $m$ and $n$. In
Appendix \ref{AppB} we additionally work out the $n=2$ case, which is special.} Probing such limits may give hints about the existence or not of
possible perturbative expansions of theories satisfying these single correlator constraints. One such example could be a theory where the order parameter field transforms in the $Z$ irrep. A similar study was performed for traceless tensors of $O(n)$ in \cite{Reehorst:2020phk} and adjoints of $SU(n)$ in \cite{Manenti:2021elk}, where theories with order parameter fields in these representations were analysed. The reader may skip this subsection on a first reading since it is rather technical, and return on a second read-through.

For $n \geq 4$ one finds
21 irreducible representations on the right-hand side of the $ n\geq 4$
$Z^{ab}_{ij} \times Z^{cd}_{kl}$ OPE.\footnote{Note that for $n=3$ there
are fewer irreps. Also, the invariant tensor $\delta_{abcdefgh}$ becomes
expressible in terms of tensors with fewer indices, see
e.g.~\cite{Kousvos:2018rhl}. We omit an analysis of the $n=3$ case in the
present work since we are interested in the large parameter limits.} We
present the projectors of the 21 irreps in Appendix \ref{zz_group_theory},
where we also explain the generalization to arbitrary group $G$. When all
indices $a$, $b$, $c$ and $d$ are different, we have three exchanged
irreps. It is convenient to define $T^{ab,cd}_{ij,kl} =
Z^{ab}_{ij}Z^{cd}_{kl}+Z^{cd}_{kl}Z^{ab}_{ij}$ and
$\bar{T}^{ab,cd}_{ij,kl}= Z^{ab}_{ij}Z^{cd}_{kl}-Z^{cd}_{kl}Z^{ab}_{ij}$,
which obviously satisfy $T^{ab,cd}_{ij,kl} = T^{cd,ab}_{kl,ij}$ and
$\bar{T}^{ab,cd}_{ij,kl} = -\bar{T}^{cd,ab}_{kl,ij}$. With this in mind, we
have
\begin{equation}
Z^{ab}_{ij} \times Z^{cd}_{kl} \sim T^{ab,cd}_{ij,kl}+\bar{T}^{ab,cd}_{ij,kl}
\,.
\end{equation}
The right-hand side may now be further decomposed by symmetrizing and
antisymmetrizing indices,
\begin{align}
\begin{split}
Z^{ab}_{ij} \times Z^{cd}_{kl} &\sim (T^{ab,cd}_{ij,kl}+T^{ac,bd}_{ik,jl}+T^{ad,bc}_{il,jk})+ (T^{ab,cd}_{ij,kl}-T^{ac,bd}_{ik,jl})+(T^{ab,cd}_{ij,kl}-T^{ad,bc}_{il,jk})\\
&\quad+(\bar{T}^{ab,cd}_{ij,kl}+
\bar{T}^{ac,bd}_{ik,jl})+(\bar{T}^{ab,cd}_{ij,kl}+\bar{T}^{ad,bc}_{il,jk})+(\bar{T}^{ab,cd}_{ij,kl}-\bar{T}^{ac,bd}_{ik,jl})+(\bar{T}^{ab,cd}_{ij,kl}-\bar{T}^{ad,bc}_{il,jk})\,,
\end{split}
\end{align}
which may also be rewritten as
\begin{align}
\begin{split}
Z^{ab}_{ij} \times Z^{cd}_{kl} &\sim \TotS^{abcd}_{ijkl}+BB^{ad,bc}_{il,jk}+ BB^{ac,bd}_{ik,jl}\\
&\quad +BZ^{ad,bc}_{il,jk}+BZ^{ac,bd}_{ik,jl}+ZB^{ad,bc}_{il,jk}+ZB^{ac,bd}_{ik,jl}
\,,
\end{split}
\end{align}
where $\TotS$ is totally symmetric in all pairs of indices,\footnote{Remember
that permuting $a$ with $b$ means we must permute $i$ with $j$, hence we do
not permute indices, but pairs of them.} whereas for $BB$ and $BZ$ we have the relations $BB^{ac,bd}_{ik,jl} = - BB^{ca,bd}_{ki,jl}= -BB^{ac,db}_{ik,lj}= BB^{bd,ac}_{jl,ik}$ and $BZ^{ac,bd}_{ik,jl} = - BZ^{ca,bd}_{ki,jl}= +BZ^{ac,db}_{ik,lj}= -BZ^{bd,ac}_{jl,ik} $. Lastly, notice that $BZ$ and $ZB$ are the same representation since $BZ^{ab,cd}_{ij,kl}=ZB^{cd,ab}_{kl,ij}$.

Next we must consider the representations where one or more of the copy (i.e. upper)
indices are equal in $Z^{ab}_{ij} \times Z^{cd}_{kl}$. These are rather
simple: if a pair of copy indices are equal we decompose onto irreps of
that copy of $G$, whereas if two copy indices are different we simply
symmetrize and antisymmetrize. Let us give some examples. Consider $a=c$
and $b\neq d$. We have
\begin{align}
Z^{ab}_{ij} \times Z^{cd}_{kl} \sim (Z^{ab}_{ij} Z^{cd}_{kl} +  Z^{ad}_{il}
Z^{cb}_{kj})+(Z^{ab}_{ij} Z^{cd}_{kl} -  Z^{ad}_{il} Z^{cb}_{kj}) \sim
RZ^{ac,bd}_{ik,jl}+RB^{ac,bd}_{ik,jl}\,,
\end{align}
where the part $R$ simply signifies that $a=c$ but is not an irrep. To
decompose $R$ onto irreps we must decompose the indices $i$ and $k$ onto
$G$ irreps and then subtract the trace of copies from the singlet of $G$:
\begin{equation}
RZ^{ac,bd}_{ik,jl} \sim
\Big(RZ^{ac,bd}_{ik,jl}+RZ^{ac,bd}_{ki,jl}-\frac{2}{m}\delta_{ik}RZ^{ac,bd}_{mm,jl}\Big)+
(RZ^{ac,bd}_{ik,jl}-RZ^{ac,bd}_{ki,jl})+\frac{2}{m}\delta_{ik}RZ^{ac,bd}_{mm,jl}\,,
\end{equation}
and
\begin{align}
\begin{split}
\delta_{ik}RZ^{ac,bd}_{mm,jl} \sim \delta_{ik}
\Big(RZ^{ac,bd}_{mm,jl}-\frac{1}{n-2}\delta_{ac}RZ^{ee,bd}_{mm,jl}\Big)+ \frac{1}{n-2}\delta_{ik}\delta_{ac}RZ^{ee,bd}_{mm,jl}
\sim XZ^{ac,bd}_{ik,jl}+ SZ^{ac,bd}_{ik,jl}\,,
\end{split}
\end{align}
where we have absorbed factors of Kronecker deltas in the representations
in order to simplify the notation.  Combining all this information together
we find
\begin{align}
\begin{split}
RZ^{ac,bd}_{ik,jl} &\sim YZ^{ac,bd}_{ik,jl}+ AZ^{ac,bd}_{ik,jl} +
XZ^{ac,bd}_{ik,jl} + SZ^{ac,bd}_{ik,jl}\,,
\end{split}
\end{align}
which we remind the reader is for $a=c$ and $b \neq d$. For the full expression one
needs to consider all possible combinations.

The last case we must consider is when we have pairs of copy indices equal,
e.g.\ $a=c$ and $b=d$. We must again decompose onto irreps of $G$. Let us
start by first decomposing $j$ and $l$ onto $G$ irreps:
\begin{equation}
Z^{ab}_{ij} \times Z^{cd}_{kl} \sim RR^{ac,bd}_{ik,jl} \sim
\Big(RR^{ac,bd}_{ik,jl}+RR^{ac,bd}_{ik,lj}-\frac{2}{m} \delta_{jl}
RR^{ac,bd}_{ik,mm}\Big)+(RR^{ac,bd}_{ik,jl}-RR^{ac,bd}_{ik,lj})+\frac{2}{m}\delta_{jl}RR^{ac,bd}_{ik,mm}\,.
\end{equation}
This is rewritten as
\begin{equation}
Z^{ab}_{ij} \times Z^{cd}_{kl} \sim
RY^{ac,bd}_{ik,jl}+RA^{ac,bd}_{ik,jl}+\frac{2}{m}\delta_{jl}RR^{ac,bd}_{ik,mm}\,,
\end{equation}
where we use $RR^{ac,bd}_{ik,jl}$ to denote an object that is not an irrep
but has $a=c$ and $b=d$. We will now focus on the last term, which is the
most complicated. We decompose the $i$ and $k$ indices onto $G$ irreps,
\begin{equation}
RR^{ac,bd}_{ik,mm}\sim \Big(RR^{ac,bd}_{ik,mm}+ RR^{ac,bd}_{ki,mm}
-\frac{2}{m}\delta_{ik} RR^{ac,bd}_{nn,mm} \Big) +(RR^{ac,bd}_{ik,mm}-
RR^{ac,bd}_{ki,mm})+ \frac{2}{m} \delta_{ik}RR^{ac,bd}_{nn,mm}\,,
\end{equation}
which is equivalent to
\begin{equation}
RR^{ac,bd}_{ik,mm}\sim YR^{ac,bd}_{ik,mm} +
AR^{ac,bd}_{ik,mm} + \frac{2}{m} \delta_{ik}RR^{ac,bd}_{nn,mm}\,.
\end{equation}
The last term is again the most complicated.\footnote{Since, for example, we know $YR^{ac,bd}_{ik,mm} \sim YR^{ac,bd}_{ik,mm}-\delta^{bd} YR^{ac,ee}_{ik,mm}/(n-1)+\delta^{bd} YR^{ac,ee}_{ik,mm}/(n-1) $ or stated in terms of the irreps $YR^{ac,bd}_{ik,mm} \sim YX^{ac,bd}_{ik,jl} + YS^{ac,bd}_{ik,jl}$ where again we have absorbed Kronecker deltas and numerical factors in the definitions of the irreps for simplicity.} Defining
$T^{ac,bd}=RR^{ac,bd}_{nn,mm}$, we subtract all possible traces to
construct irreducible objects:
\begin{align}
\begin{split}
T^{ac,bd}&\sim T^{ac,bd}-\frac{1}{n-2}(\delta_{bd}T^{ac,ee}+ \delta_{ac}T^{ee,bd})+\frac{\delta_{ac}\delta_{bd}}{(n-1)(n-2)}T^{ee,ff}\\
&\quad+\frac{\delta_{bd}}{n-2}\Big(T^{ac,ee}-\frac{\delta_{ac}}{n}T^{ee,ff}\Big)+\frac{\delta_{ac}}{n-2}\Big(T^{ee,bd}-\frac{\delta_{bd}}{n}T^{ee,ff}\Big)\\
&\quad-\delta_{ac}\delta_{bd}\Big(\frac{1}{(n-1)(n-2)}-\frac{2}{n(n-2)}
\Big)T^{ee,ff}\\
&\sim \bar{X}^{ac,bd} + XS^{ac,bd}+ SX^{ac,bd} + \delta_{ac}\delta_{bd}S\,,
\end{split}
\end{align}
where again we have absorbed various factors in the definitions of the
irreps. The other irreps follow rather straightforwardly, hence we omit a
detailed analysis. In conclusion we find the 21 irreps ($S$, $XS$,
$\bar{X}$, $XY$, $XA$, $SY$, $SA$, $YY$, $AA$, $YA$, $BB$, $\TotS$, $XB$,
$XZ$, $SB$, $SZ$, $YB$, $YZ$, $AB$, $AZ$, $BZ$) with respective dimensions
($S:1$, $XS:n-1$, $\bar{X}:n(n-3)/2$, $XY:n(n-2)(m-1)(m+2)/2$,
$XA:n(n-2)(m-1)m/2$, $SY:n(m-1)(m+2)/2$, $SA:n(m-1)m/2$,
$YY:n(n-1)(m+2)^2(m-1)^2/8$, $AA:n(n-1)(m-1)^2 m^2 /8 $, $YA:n(n-1)(m-1)^2
m (m+2)/4$, $ BB:2m^4n(n-1)(n-2)(n-3)/(4!) $, $\TotS:m^4n(n-1)(n-2)(n-3)/(4!)$,
$XB:(n-3)m^2 n(n-1)/2$, $XZ:(n-3)m^2 n(n-1)/2$, $SB:m^2n(n-1)/2$,
$SZ:m^2n(n-1)/2$, $YB:(m+2)(m-1)m^2n^2(n-1)/4$,
$YZ:(m+2)(m-1)m^2n^2(n-1)/4$, $AB:(m-1)m^3n^2(n-1)/4$,
$AZ:(m-1)m^3n^2(n-1)/4$, $BZ:3m^4n(n-1)(n-2)(n-3)/(4!)$).

\newsec{Numerical results}[numerics]
With the group theory outlined above and the sum rules given in the appendices we are ready to start obtaining numerical results. We start by probing the system of four point correlators obtained by considering all combinations of external operators $\phi^a_i$ and $X^{bc}$. We then conclude the chapter by presenting bounds pertaining to four point functions of symmetric tensors $Z^{ab}_{ij}$.
\subsec{Islands close to the unitarity bound}
In this section we study the portion of parameter space ``close'' to the
unitarity bound. By this we refer to values of the order parameter scaling
dimensions roughly around $0.5\text{-}0.53$; these correspond to typical
values found in fixed points of multi-scalar theories. However, this does
not necessarily mean that the theories we find are due to the fixed point
of some multi-scalar Lagrangian description. For $n$ sufficiently large we
find kinks, and corresponding islands, that converge to the expected
values, that is $\Delta_\phi = \Delta_\phi^{O(m)}$ and $\Delta_X =
\Delta_S^{O(m)}$.\footnote{Whereas for the scalar singlet one has and
$\Delta_S = d - \Delta_S^{O(m)}$, although we do not present results for
this in the present work.} Here $\Delta_\phi^{O(m)}$ is the scaling dimension of the order parameter field  and $\Delta_S^{O(m)}$ is the scaling dimension of the lowest lying scalar singlet, both in the theory of $n$ decoupled $O(m)$ models. For smaller values
of $n$, which also correspond to the phenomenologically interesting values,
it is not clear whether the kinks (and their islands) correspond to the
ordinary $d=4-\varepsilon$ fixed points or are some new hitherto unknown
CFTs.

We separate our results into two groups depending on how we obtain the
islands. In the case where the islands are found in the
$\Delta_\phi$-$\Delta_X$ plane of
parameter space, we impose gaps on certain sectors mainly guided by the
extremal functional method \cite{El-Showk:2012vjm,
El-Showk:2016mxr}.\footnote{One can also try to justify the gaps from the
large $n$ point of view. For example, we know that the leading $X^{ab}$
operator has scaling dimension $\Delta_X = \Delta_S^{O(m)}+O(1/n)$, so we expect the subleading operator to have dimension either $\Delta_{X^\prime}= \Delta_{S^\prime}^{O(m)}+O(1/n)$ or $\Delta_{X^\prime}= 2\Delta_{S}^{O(m)}+O(1/n)$ or $\Delta_{X^\prime}= \Delta_{S}^{O(m)}+2+O(1/n)$ (remember $X^{ab}\sim (\delta^{abcd}-\frac{1}{n}\delta^{ab}\delta^{cd})\phi^c_i \phi^d_i$ in the weakly coupled limit). All these satisfy $\Delta_{X^\prime}>3$ at $n \rightarrow \infty$. The main issue with this line of reasoning though is that we do not have the explicit corrections to subleading order in $1/n$.}
Since these gaps are motivated by the extremal functional, their choice is
not rigorous. Nevertheless, we believe that even in absence of rigor our
islands are sufficient to motivate that something particularly interesting
takes place in the corresponding region of parameter space.  We also
present islands in the $\Delta_\phi$-$\Delta_S$ plane, specifically in the
phenomenologically interesting cases $O(2)^2 \rtimes S_2$ and $O(2)^3
\rtimes S_3$. For these cases, in addition to assumptions described above,
we demand the saturation of certain exclusion bounds. More specifically, we
demand that the exclusion bounds in the $X$ sector\footnote{These are the
exclusion plots that display kinks/changes of slope.} are saturated.

Obtaining islands by demanding saturation of exclusion bounds (in our case
the $X$ bound) is somewhat morally similar to the extremal functional
method where one may again demand saturation of a bound to then find an
approximate spectrum that corresponds to it. However, we believe that our
method minimizes the probability that a zero of the extremal function is
spurious. This is because in our approach we do not try to find approximate
zeroes of the functional, but instead show that the positions of these
zeroes cannot be excluded by the bootstrap algorithm (whereas the parameter
space surrounding them can be). Also, in conjunction with additional
assumptions we can provide a minimum and maximum value for $\Delta_\phi$.
Lastly, we note that we have found these islands to depend smoothly on the
precise position of the $X$ bound. In other words, if $\Delta_X$ were to
change a little there would be no major changes in the corresponding
$\Delta_\phi$-$\Delta_S$ plane island. We confirmed this behavior while
working on \cite{Kousvos:2019hgc}, albeit for a slightly different
symmetry, namely $\mathbb{Z}_2^{\;\,n}\rtimes S_n$. Another way to think of our method for obtaining islands in the $\Delta_\phi$-$\Delta_S$ plane, is that we apply the usual bootstrap algorithm, but to a specific slice of parameter space. This slice is precisely the one that maximizes the scaling dimension $\Delta_X$.

We note that for our numerical calculations in the present work we use
\texttt{qboot} \cite{Go:2020ahx}\footnote{We also encourage the reader to
see \cite{Go:2019lke}, which automates the derivation of sum rules,
although we did not use it for the present paper since our groups of
interest were not supported.} and \texttt{SDPB}
\cite{Simmons-Duffin:2015qma, Landry:2019qug}.  Some of the spectrum
calculations were performed with \texttt{PyCFTBoot} \cite{Behan:2016dtz}.

\begin{figure}[H]
  \centering
  \includegraphics[scale=0.9]{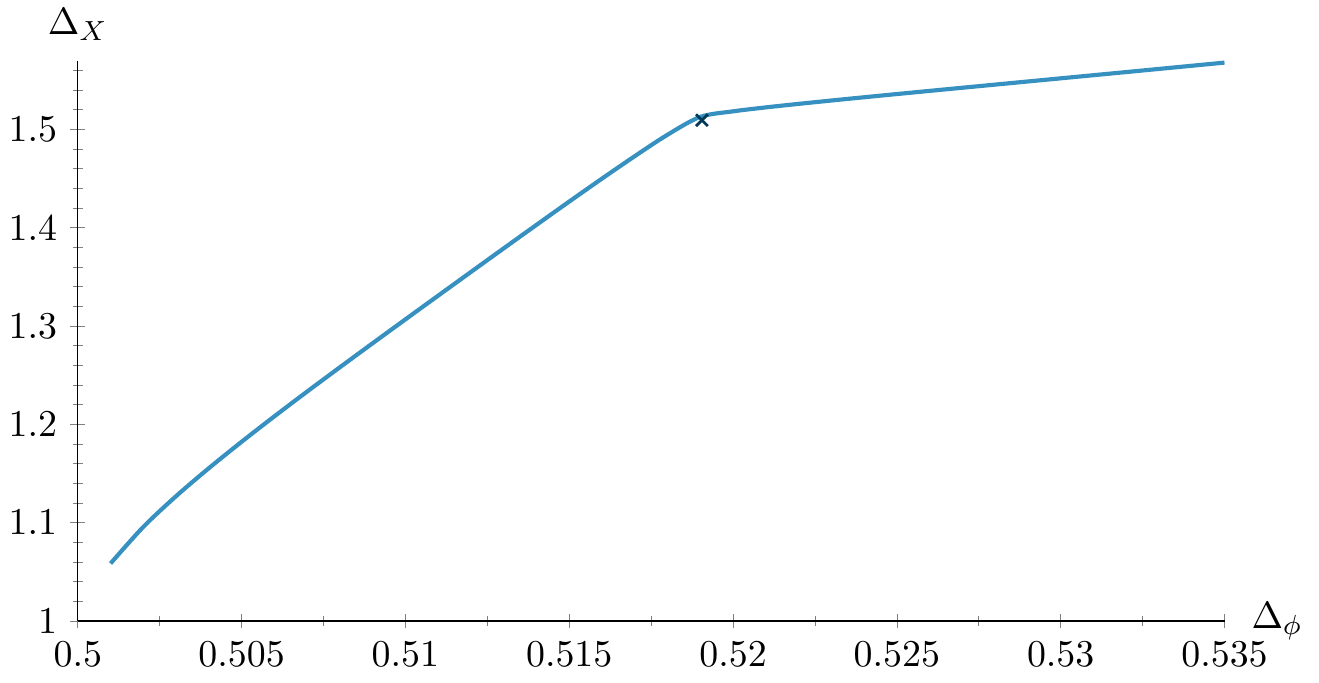}
  \caption{Single correlator $\MNmath_{2, 100}$ exclusion bound. The cross
  is the $O(2)$ model \cite{Chester:2019ifh}. The line corresponds to the maximum allowed scaling
  dimension of the first scalar $X$ operator as a function of the scaling
  dimension of the order parameter operator. In \texttt{qboot} we used
  $\Lambda = 45$, $\ell=\{0,\ldots, 50, 55, 56, 59, 60, 64, 65, 69, 70, 74,
  75, 79, 80, 84, 85, 89, 90\}$ and $\nu_{\max}=20$.}
  \label{FigMN2100}
\end{figure}

\begin{figure}[H]
  \centering
  \includegraphics[scale=1]{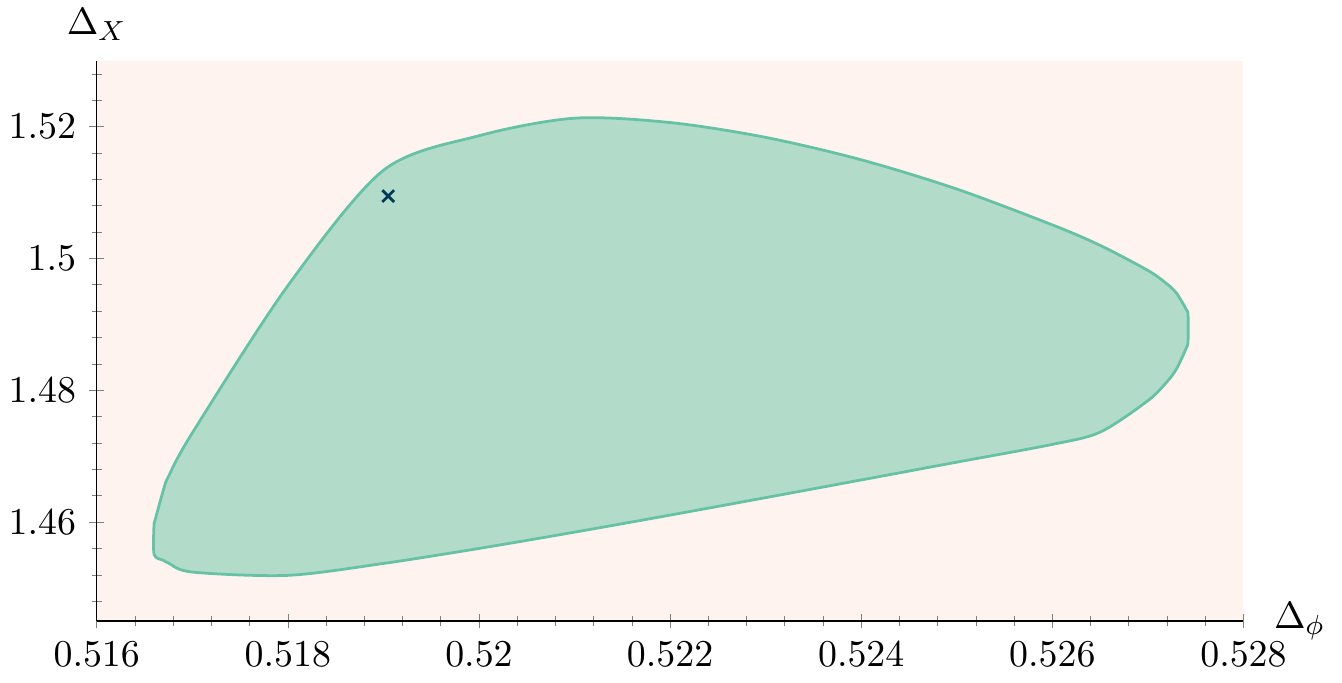}
  \caption{$\MNmath_{2, 100}$ island corresponding to the kink in Fig.\
  \ref{FigMN2100} obtained using the mixed $\phi\text{-}X$ system of
  correlators. The cross is the $O(2)$ model \cite{Chester:2019ifh}. The island assumes that the
  second scalar $X$ and spin-1 $A$ operators have scaling dimensions that
  satisfy $\Delta \geq 3.0$, and the first spin-$2$ singlet after the
  stress tensor has a dimension that satisfies $\Delta \geq 4.0$. The first
  spin-1 $A$ operator satisfies $\Delta_A =2.0$ since it is the conserved
  vector of $O(2)$, and the first spin-2 $S$ operator satisfies
  $\Delta_{T_{\mu \nu}}=3.0$. We also fixed the ratio of the OPE
  coefficients with which the stress tensor appears, $\lambda_{\phi \phi
  T_{\mu \nu}}/\lambda_{XX T_{\mu \nu}} = \Delta_\phi / \Delta_X$. Lastly,
  we imposed $\Delta_{\phi^\prime} \geq 1.0$ for the second operator in the
  defining representation. In \texttt{qboot} we used $\Lambda = 27$,
  $\ell=\{0,\ldots,50, 55, 56, 59, 60, 64, 65, 69, 70, 74, 75, 79, 80, 84,
  85, 89, 90\}$ and $\nu_{\max}=25$.}
  \label{FigMN2100island}
\end{figure}

\begin{figure}[H]
  \centering
  \includegraphics[scale=0.9]{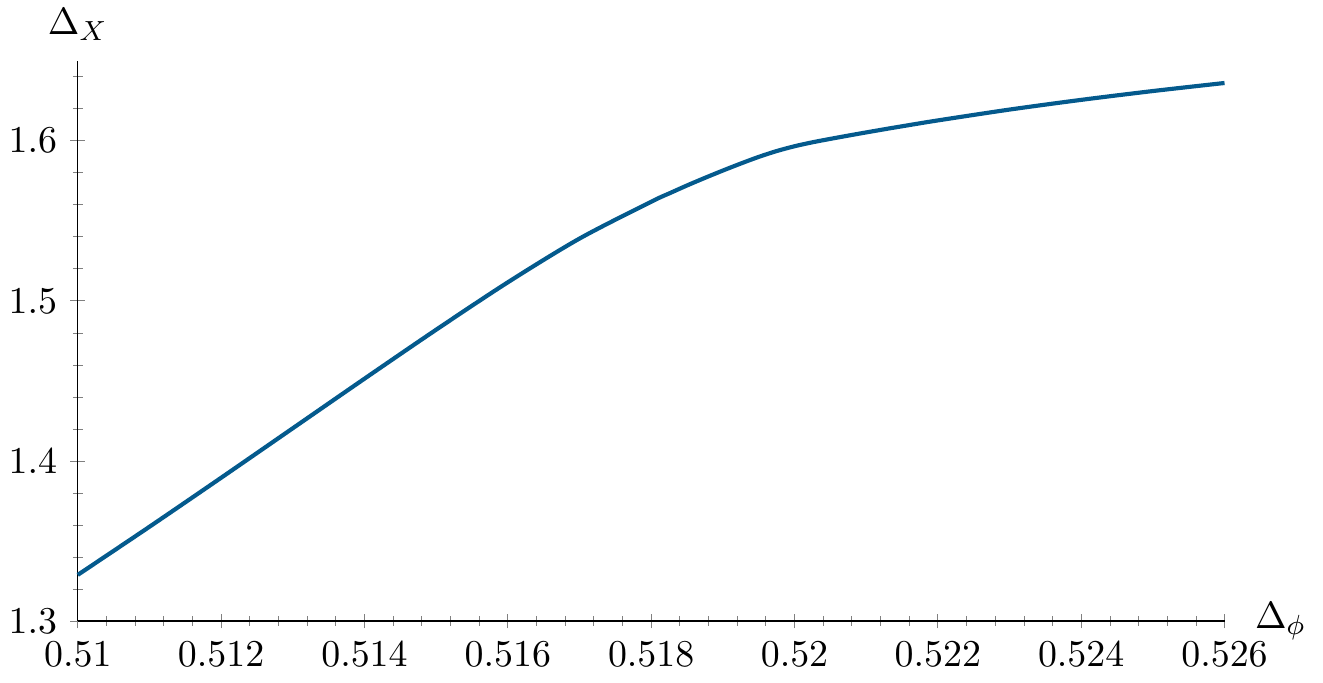}
  \caption{Single-correlator $\MNmath_{2, 3}$ exclusion bound. The line
  corresponds to the maximum allowed scaling dimension of the first scalar
  $X$ operator as a function of the scaling dimension of the order
  parameter operator. In \texttt{qboot} we used $\Lambda = 45$,
  $\ell=\{0,\ldots, 50, 55, 56, 59, 60, 64, 65, 69, 70, 74, 75, 79, 80, 84,
  85, 89, 90\}$ and $\nu_{\max}=20$.}
  \label{FigMN23}
\end{figure}

\begin{figure}[H]
  \centering
  \includegraphics[scale=1]{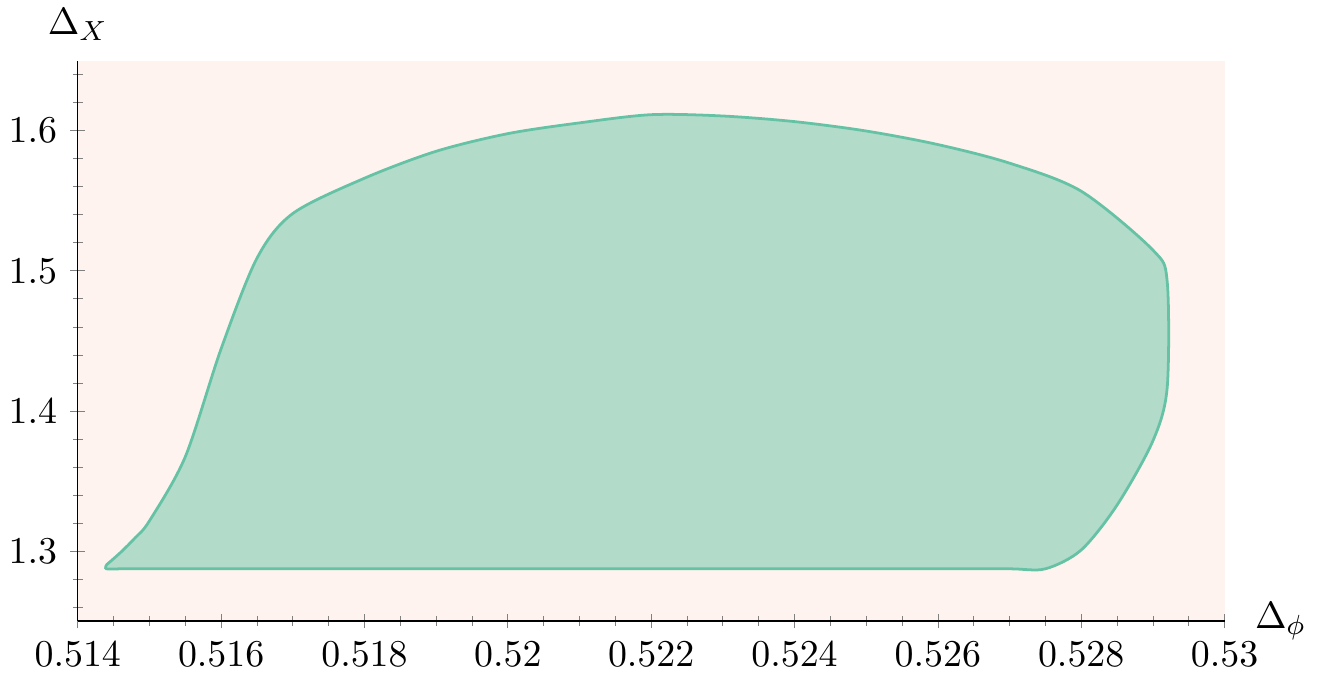}
  \caption{$\MNmath_{2, 3}$ island corresponding to the kink in Fig.\
  \ref{FigMN23} obtained using the mixed $\phi$-$X$ system of correlators.
  The island assumes that the second scalar $X$ and spin-1 $A$ operators
  have scaling dimensions that satisfy $\Delta \geq 3.0$ and the first
  spin-$2$ singlet after the stress tensor has a dimension that satisfies
  $\Delta \geq 4.0$. The first spin-1 $A$ operator satisfies $\Delta_A
  =2.0$ since it is the conserved vector of $O(2)$, and the first spin-$2$
  $S$ operator satisfies $\Delta_{T_{\mu \nu}}=3.0$. We also fixed the
  ratio of the OPE coefficients with which the stress tensor appears,
  $\lambda_{\phi \phi T_{\mu \nu}}/\lambda_{XX T_{\mu \nu}} = \Delta_\phi /
  \Delta_X$.  Lastly, we imposed $\Delta_{\phi^\prime} \geq 1.0$ for the
  next-to-leading operator in the defining representation. To obtain this
  figure with \texttt{qboot} \cite{Go:2020ahx} we used $\Lambda = 27$,
  $\ell=\{0,\ldots,50, 55, 56, 59, 60, 64, 65, 69, 70, 74, 75, 79, 80, 84,
  85, 89, 90\}$ and $\nu_{\max}=25$.}
  \label{FigMN23Xisland}
\end{figure}

\subsubsec{Islands in the
  \texorpdfstring{$\Delta_\phi\text{-}\Delta_X$}{Delta\_phi-Delta\_X}  plane of
parameter space}\label{phi-X}
In Figs.\ \ref{FigMN2100island} and \ref{FigMN23Xisland} we plot the
islands corresponding respectively to the kinks in Figs.\ \ref{FigMN2100}
and \ref{FigMN23}. Note that we do not present an island in the
$\Delta_\phi$-$\Delta_X$ plane for $n=2$, even though it is phenomenologically interesting. This is because up to $\Lambda =60$ in \texttt{qboot} the allowed region obtained is very large. For Fig.\ \ref{FigMN2100island} and Fig.\
\ref{FigMN2100} we see good agreement with the perturbative
expectation.\footnote{SRK thanks Bernardo Zan for pointing out aspects of
the large $n$ description.} That is, as $n \rightarrow \infty$ we expect
$\Delta_X \rightarrow {\Delta_S}^{O(m)}$, hence for Fig.\
\ref{FigMN2100island} in particular we have $\Delta_X \sim
{\Delta_S}^{O(2)}=1.50946(22)$ and  $\Delta_\phi \sim
{\Delta_\phi}^{O(2)}=0.519050(40)$ \cite{Chester:2019ifh}. The interested
reader is referred to \cite{PhysRevLett.31.1494} for the large $n$ limit in
the case $G=\mathbb{Z}_2$ or to \cite{Binder:2021vep} for a more recent
reference.  For Figs.\ \ref{FigMN23Xisland} and \ref{FigMN23} the situation
is more complicated. We cannot make any conclusive statements with regards
to the theory captured by this island.

\subsubsec{Islands in the
  \texorpdfstring{$\Delta_\phi\text{-}\Delta_S$}{Delta\_phi-Delta\_S}
plane of parameter space} The motivation behind studying the islands in the
$\Delta_\phi$-$\Delta_S$ plane of parameter space is twofold. On the one
hand, the $\Delta_\phi$-$\Delta_S$ plane of parameter space is the most
immediately relevant one in terms of critical exponents since $\nu
=1/(3-\Delta_S)$ and $\beta =\Delta_\phi / (3-\Delta_S)$. On the other, this way
(i.e.\ by demanding saturation of the $X$ sector exclusion bound) we can
study specifically the theory that creates the kink by excluding the rest
of parameter space. We do again emphasize that this approach is not
rigorous in the usual sense of the term as used in the bootstrap.

\begin{figure}[H]
  \centering
  \includegraphics[scale=1]{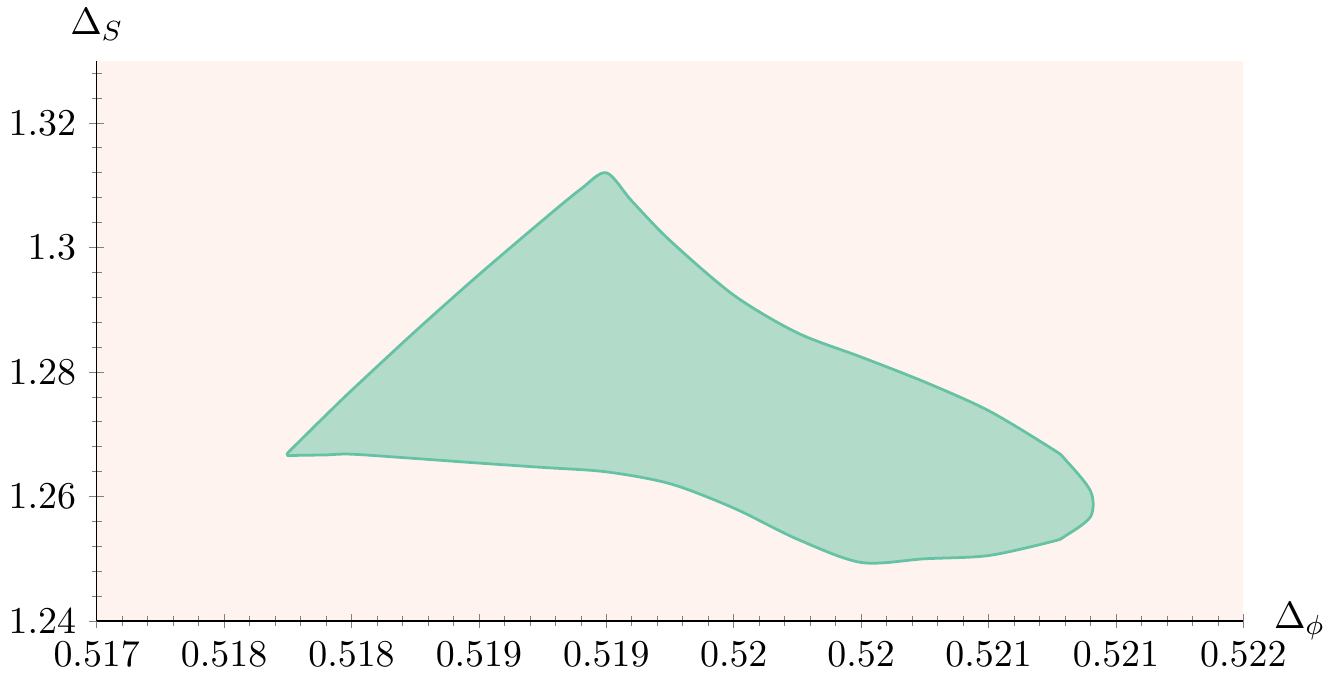}
  \caption{$\MNmath_{2, 3}$ island corresponding to the theory saturating
  Fig.\ \ref{FigMN23} obtained using the mixed $\phi\text{-}X$ system of
  correlators. To obtain the island we imposed $\Delta_{S^\prime} \geq
  3.0$, $\Delta_A =2.0$ ($O(2)$ conserved vector), $\Delta_{A^\prime} \geq
  3.0$, $\Delta_{X^\prime} \geq 3.0$ and $\Delta_{\phi^\prime} \geq 1.0$.
  Lastly, we assumed that the first scalar $X$ operator saturates the bound
  of Fig.\ \ref{FigMN23}. To obtain this figure with \texttt{qboot} we used
  $\Lambda = 27$, $\ell=\{0,\ldots,50, 55, 56, 59, 60, 64, 65, 69, 70, 74,
  75, 79, 80, 84, 85, 89, 90\}$ and $\nu_{\max}=25$.}
  \label{FigMN23island}
\end{figure}

\begin{figure}[H]
  \centering
  \includegraphics[scale=1]{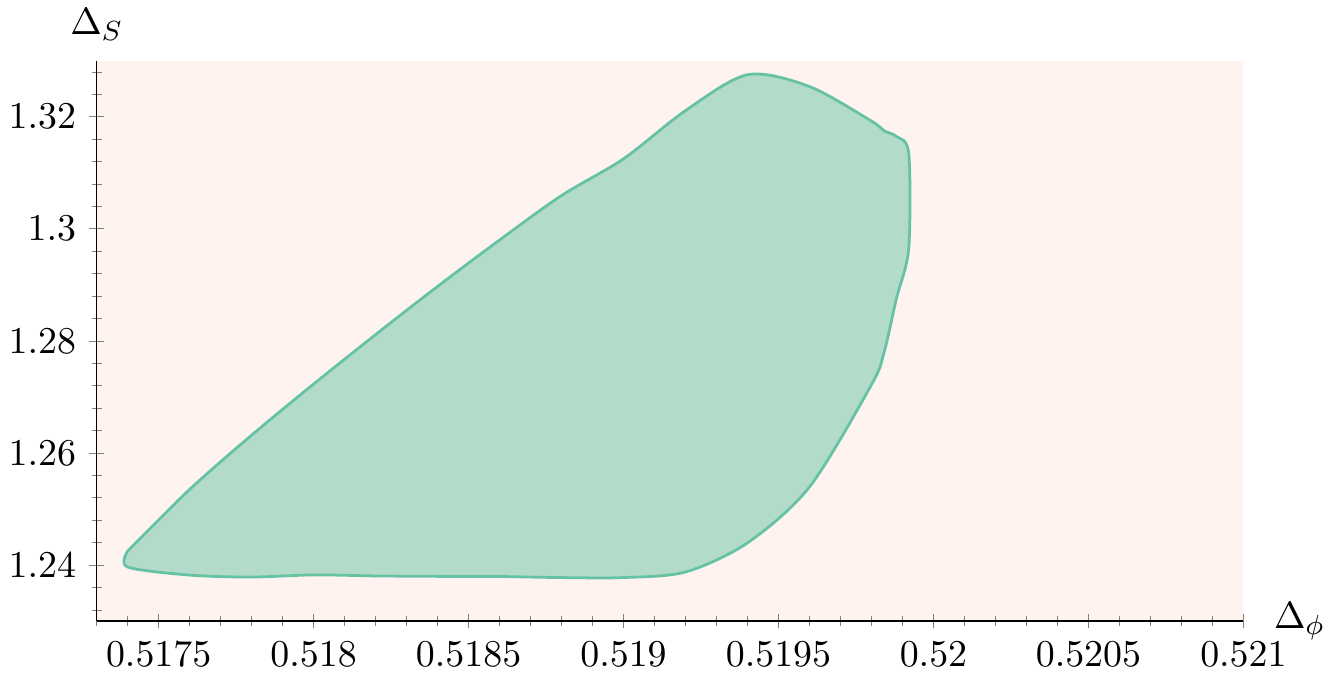}
  \caption{$\MNmath_{2, 3}$ island corresponding to the theory saturating
  Fig.\ \ref{FigMN23} obtained using the mixed $\phi\text{-}X$ system of
  correlators.  To obtain the island we imposed $\Delta_{S^\prime} \geq
  3.5$, $\Delta_A =2.0$ ($O(2)$ conserved vector), $\Delta_{A^\prime} \geq
  3.9$, $\Delta_{X^\prime} \geq 3.0$ and $\Delta_{\phi^\prime} \geq 1.5$.
  Lastly, we assumed that the first $X$ operator saturates the bound of
  Fig.\ \ref{FigMN23}. To obtain this figure with \texttt{qboot} we used
  $\Lambda = 20$, $\ell=\{0,\ldots,50, 55, 56, 59, 60, 64, 65, 69, 70, 74,
  75, 79, 80, 84, 85, 89, 90\}$ and $\nu_{\max}=25$.}
  \label{FigMN23islandshrinked}
\end{figure}

\begin{figure}[H]
  \centering
  \includegraphics[scale=1]{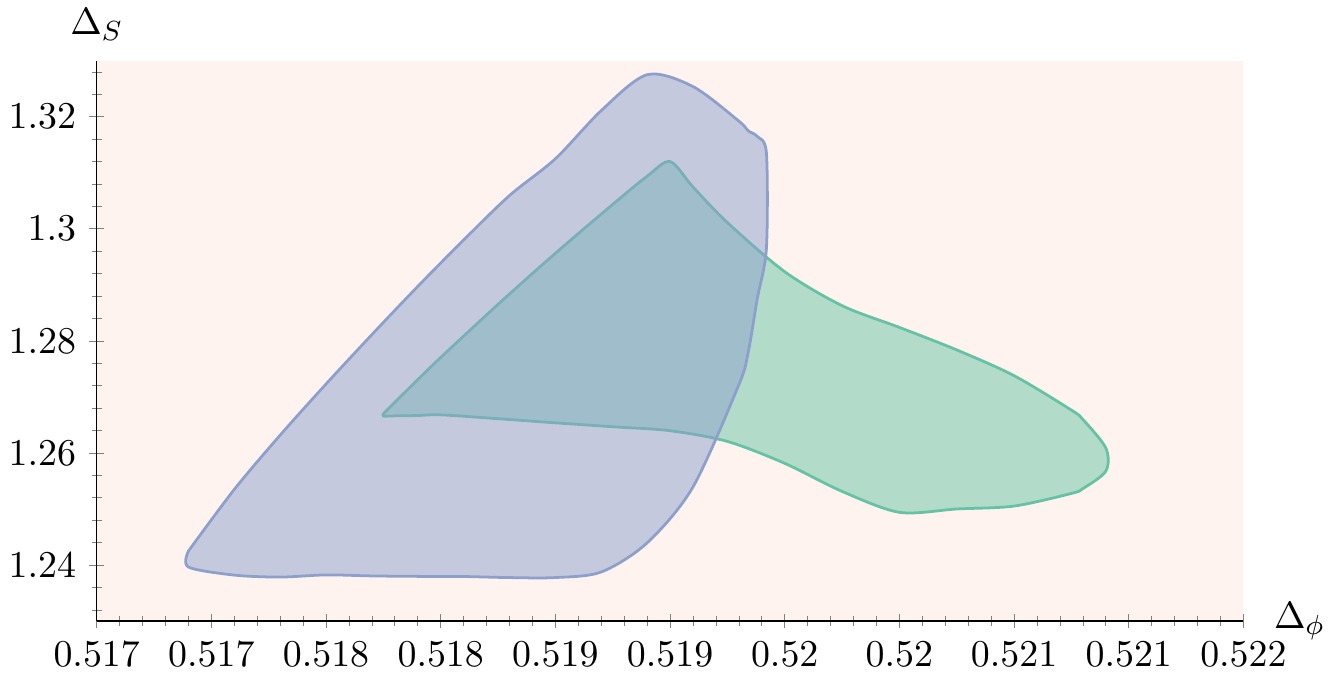}
  \caption{$\MNmath_{2, 3}$ islands of the last two figures, namely Fig.\
    \ref{FigMN23island} and Fig.\ \ref{FigMN23islandshrinked}, in a
    combined plot. We emphasize that the blue island is computed at a lower
    $\Lambda$, and that is why it appears larger on the left despite the
    stronger spectrum assumptions used to obtain it compared to the green
    island. The stronger spectrum assumptions have an effect on the right
    part of the island.}
  \label{FigMN23islandComb}
\end{figure}

With these considerations in mind, in Figs.\ \ref{FigMN23island} and
\ref{FigMN22island} we obtain islands pertaining to the theories saturating
the corresponding $\MNmath_{{2,3}}$ and $\MNmath_{{2,2}}$ $X$ sector
bounds. Note that in Fig.\ \ref{FigMN23islandshrinked} we present again the
$MN_{2,3}$ island but with much larger gaps in order to demonstrate that as
our gaps approach the values predicted by the extremal functional method
the islands become more smooth. Lastly, in Fig.\ \ref{FigMN23islandComb} we explicitly show the overlap of these two figures.

\begin{figure}[H]
  \centering
  \includegraphics[scale=1]{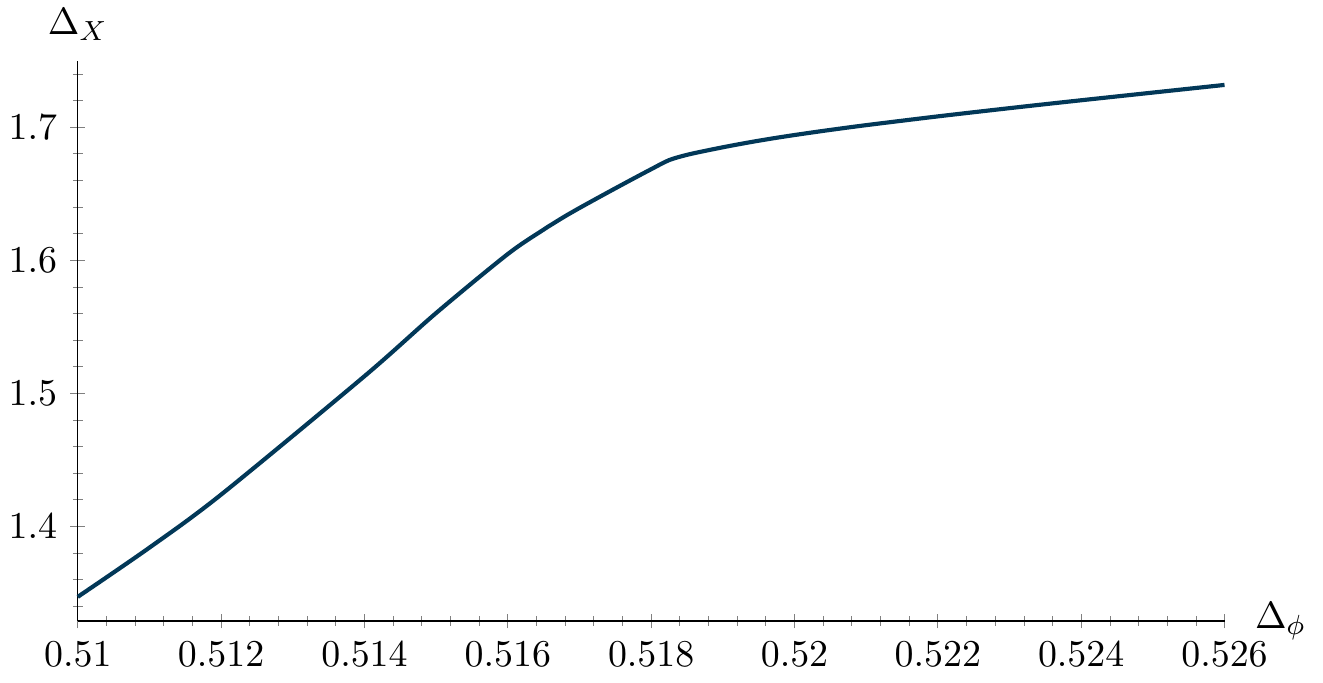}
  \caption{Mixed correlator ($\phi\text{-}X$) $\MNmath_{2, 2}$ exclusion
  bound. The line corresponds to the maximum allowed scaling dimension of
  the first scalar $X$ operator as a function of the scaling dimension of
  the order parameter operator. Here we used the mixed- instead of the
  single-correlator system, since in this particular case the sum rules are
  simpler and hence numerically cheaper, than the $n \geq 3$ cases. To
  obtain this figure with \texttt{qboot} we used $\Lambda = 45$,
  $\ell=\{0,\ldots,50, 55, 56, 59, 60, 64, 65, 69, 70, 74, 75, 79, 80, 84,
  85, 89, 90\}$ and $\nu_{\max}=20$. }
  \label{FigMN22}
\end{figure}

\begin{figure}[H]
  \centering
  \includegraphics[scale=1]{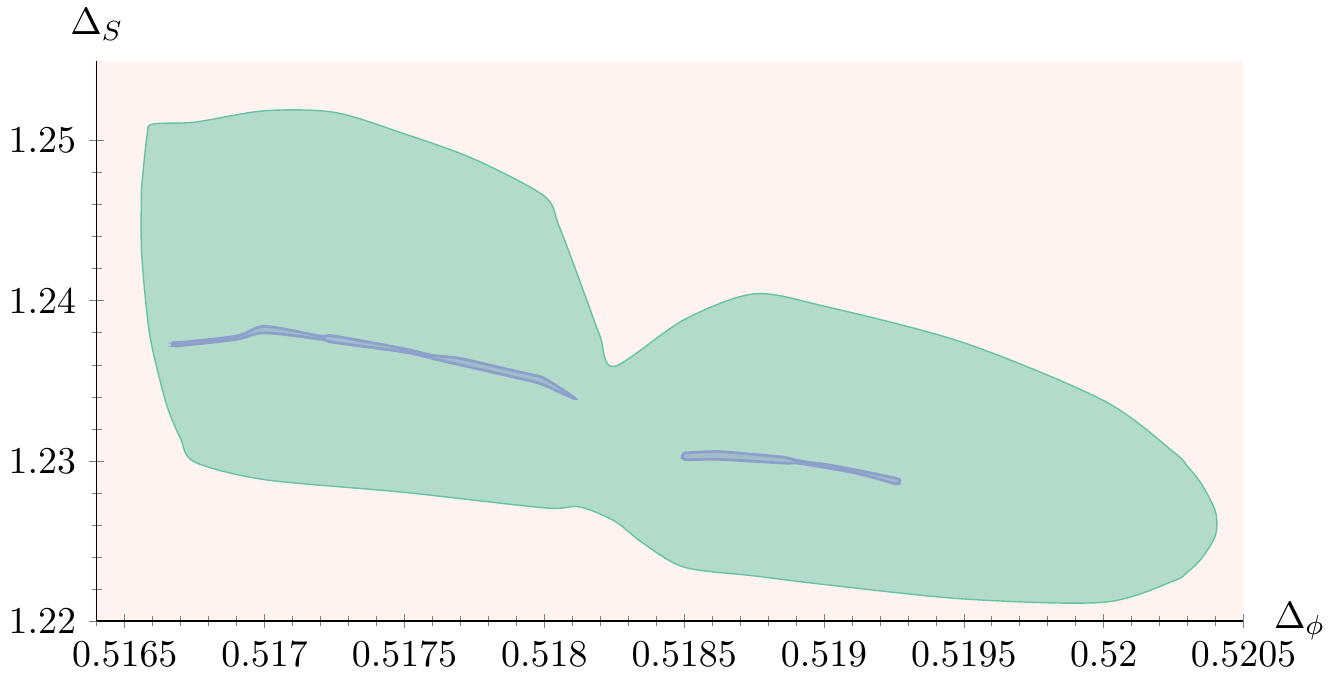}
  \caption{$\MNmath_{2, 2}$ island corresponding to the theory saturating
  Fig.\ \ref{FigMN22} obtained using the mixed $\phi$-$X$ system of
  correlators.  To obtain the island we imposed $\Delta_{S^\prime} \geq
  3.0$, $\Delta_A =2.0$ ($O(2)$ conserved vector), $\Delta_{A^\prime } \geq
  3.0$, $\Delta_{X^\prime } \geq 3.0$ and $\Delta_{\phi^\prime} \geq 1.0$.
  Lastly, we assumed that the first scalar $X$ operator saturates the bound
  of Fig.\ \ref{FigMN22}. In \texttt{qboot} we used $\Lambda = 35$,
  $\ell=\{0,\ldots, 50, 55, 56, 59, 60, 64, 65, 69, 70, 74, 75, 79, 80, 84,
  85, 89, 90\}$ and $\nu_{\max}=25$. We also display (in blue) two narrow
  islands that correspond to the allowed region that remains if we choose
  to saturate the $\Delta_X$ bound at $\Lambda = 35$ instead of $\Lambda
  =45$ (Fig.\ \ref{FigMN22} is obtained at $\Lambda =45$).}
  \label{FigMN22island}
\end{figure}

Additionally, note that the rightmost tip of the left blue island in Fig.\ \ref{FigMN22island} extends to the $\Delta_\phi$ of the kink of Fig.\ \ref{FigMN22}. Thus, the putative theory that lives at that kink is allowed under the assumptions mentioned in the caption of Fig.\ \ref{FigMN22island}. Given that this is a marginal case, however, further numerical work with stronger numerics and more refined methods is required to provide further clarity.

\begin{figure}[H]
  \centering
  \includegraphics[scale=1]{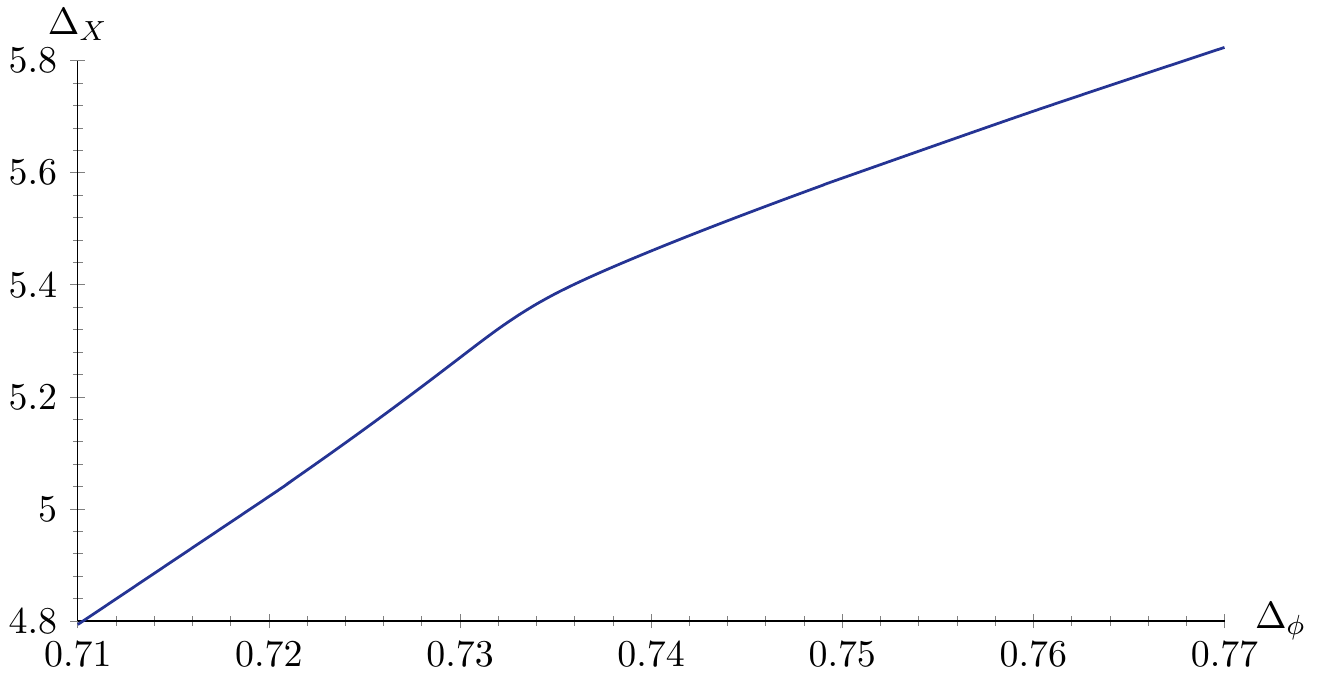}
  \caption{Single-correlator $\MNmath_{20,2}$ exclusion bound. The line
  corresponds to the maximum allowed scaling dimension of the first scalar
  $X$ operator as a function of the scaling dimension of the order
  parameter operator. In \texttt{qboot} we used $\Lambda = 45$,
  $\ell=\{0,\ldots, 60, 63, 64, 66, 67, 73, 74, 77, 78, 81, 82, 85, 86, 89,
  90, 93, 94, 97, 98\}$ and $\nu_{\max}=26$.}
  \label{X_20_2_bound}
\end{figure}

\subsec{Islands further away from the unitarity bound}
In \cite{Stergiou:2019dcv} more kinks further away from the unitarity bound
were observed, such as the one in Fig.\ \ref{X_20_2_bound}.  These were subsequently studied in \cite{Henriksson:2021lwn}. In this work
we probe them using two different strategies. The first strategy is to
bootstrap them directly, i.e.\ to demand that the exclusion plot with the
kink is saturated and impose assumptions inspired by the extremal
functional method to obtain an isolated allowed region. The second strategy
is to try and study the kinks indirectly, based on the observation made in
\cite{Henriksson:2021lwn} that the spectrum at these kinks contained
operators in other representations very close to the unitarity bound,
namely fields in the $Y$ and $Z$ irreps.  Hence, since the numerical
bootstrap tends to be stronger closer to the unitarity bound, we can
bootstrap these fields instead of the initial $\phi$ field we were
studying. We only give a brief example of this strategy in this work, and
leave a more complete treatment to future work.

An interesting observation about the second kinks is that, to our
knowledge, they have appeared in all works studying scalars involving the
group $S_n$. More specifically they have appeared in \cite{Rong:2017cow}
($S_n\times\mathbb{Z}_2$), \cite{Kousvos:2021mpw}
($\mathbb{Z}_2^{\;\,n}\rtimes S_n$) and \cite{Henriksson:2021lwn} ($O(m)^n
\rtimes S_n$). Notably, they always appear in what we call the $X$ sector
in this work. Second kinks have also appeared in other theories
\cite{Li:2018lyb, Li:2020bnb, He:2020azu, Reehorst:2020phk, He:2021xvg,
Manenti:2021elk}.

Note that in Fig.\ \ref{MN_20_2_island} we were still able to obtain an
island even if we imposed $\Delta_{S^\prime} \geq 5.0$. This is not
allowed, though, for unitary CFTs in 3D, as can be seen from \cite[Fig.\
3]{Nakayama:2016jhq}.\footnote{We thank Alessandro Vichi for bringing this
to our attention} In other words, the bootstrap seems to be completely
insensitive to that operator. This is not the first time we observe this
phenomenon. For example when working on \cite{Henriksson:2020fqi}, where we
had concrete large $n$ predictions to compare to, the bootstrap seemed to
be completely insensitive to the second scalar singlet at the antichiral
fixed point for $m=2$ and $n=10$, even though all other CFT data agreed
exceptionally well with the perturbative predictions. We believe that
sensitivity to the second scalar singlet should be restored once $S$ is
included as an external operator.

\begin{figure}[H]
  \centering
  \includegraphics[scale=1]{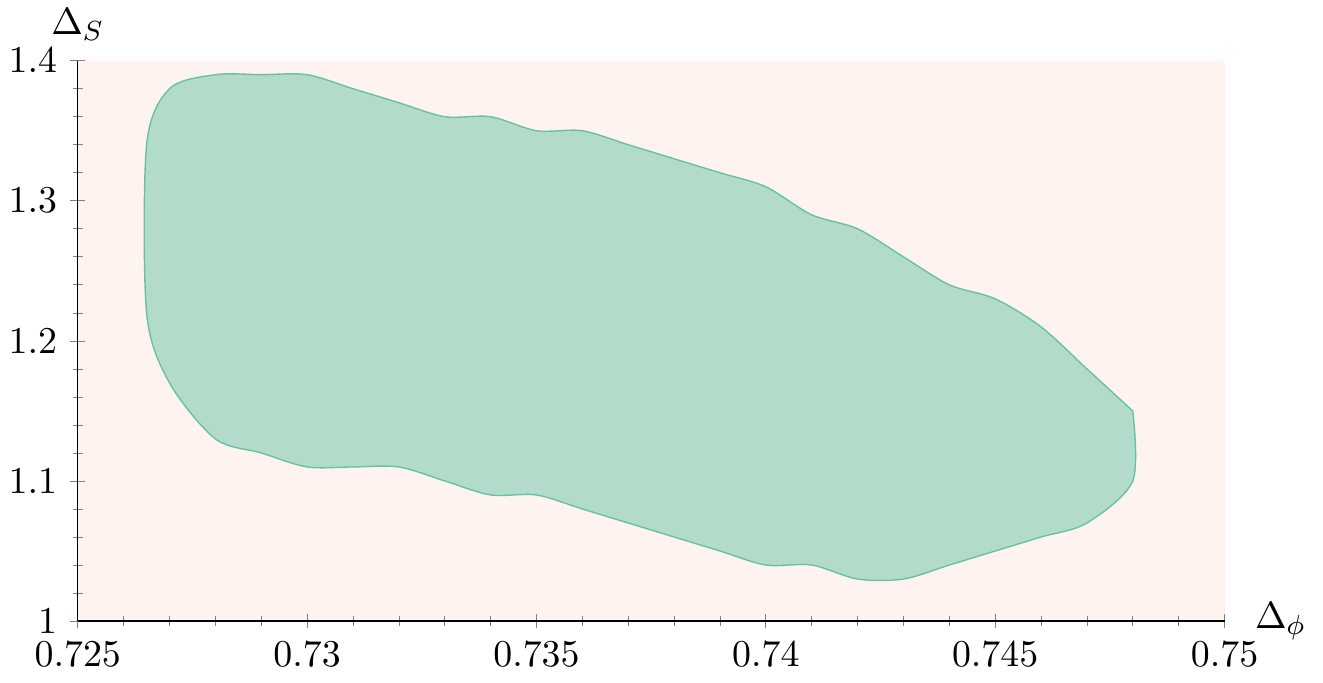}
  \caption{$\MNmath_{20, 2}$ island in the $\phi$-$S$ plane of parameter
  space. To obtain this figure we assumed that the first spin-two singlet
  is the stress-energy tensor and that $X$ saturates the bound of Fig.\
  \ref{X_20_2_bound}. We further imposed $\Delta_{S'} \geq 3.0$,
  $\Delta_{T_{\mu \nu}^\prime}  \geq 6.0$ (next-to-leading spin-two
  singlet), $\Delta_{X'} \geq 10.0$, $\Delta_{\phi^\prime} \geq1.0$. In
  \texttt{qboot} we used $\Lambda = 35$, $\ell=\{0,\ldots, 50, 55, 56, 59,
  60, 64, 65, 69, 70, 74, 75, 79, 80, 84, 85, 89, 90\}$ and
  $\nu_{\max}=25$.}
  \label{MN_20_2_island}
\end{figure}

\begin{figure}[H]
  \centering
  \includegraphics[scale=1]{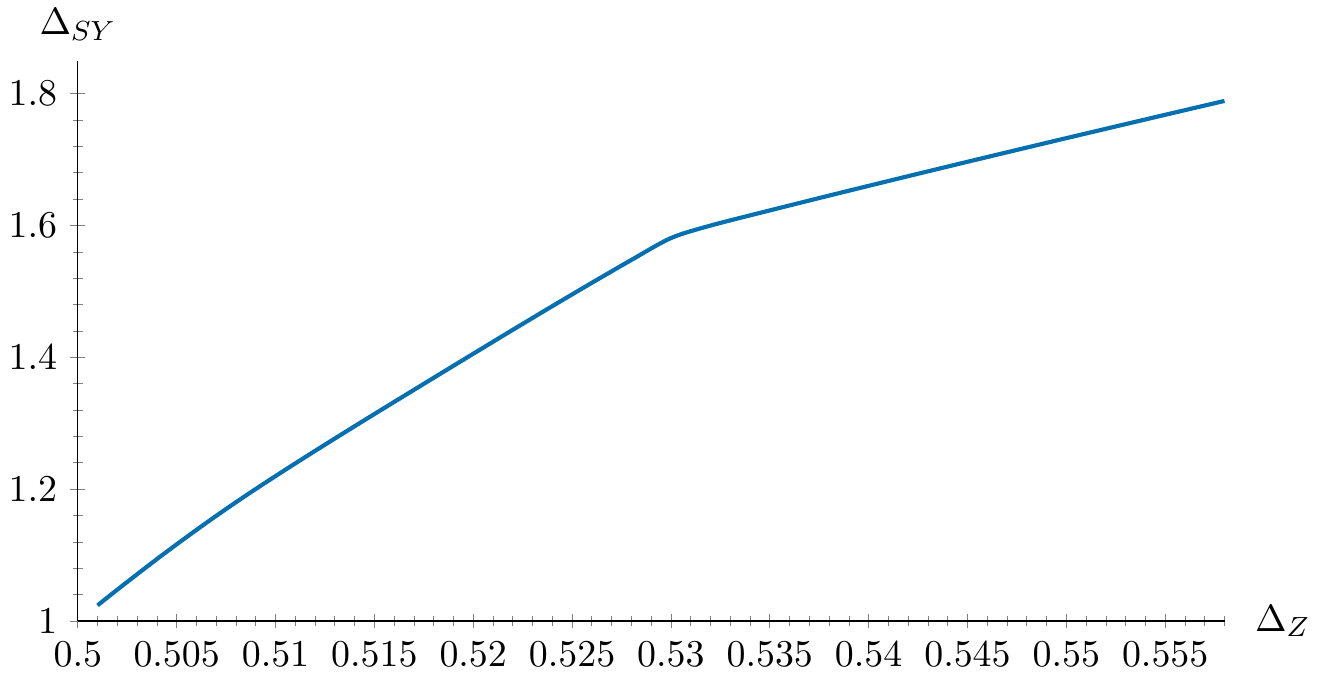}
  \caption{Single correlator $\MNmath_{100, 2}$ exclusion bound using the
  $\langle ZZZZ \rangle$ crossing equation. The line corresponds to the
  maximum allowed scaling dimension of the first $SY$ (remember $SY=Y$)
  operator as a function of the scaling dimension of the order parameter
  operator. To obtain this figure with \texttt{qboot} we used $\Lambda =
  20$, $\ell=\{0, \ldots, 50, 55, 56, 59, 60, 64, 65, 69, 70, 74, 75, 79,
  80, 84, 85, 89, 90\}$ and $\nu_{\max}=25$.}
  \label{ZZ_100_2}
\end{figure}

\subsec{The \texorpdfstring{$n=2$}{n=2} \texorpdfstring{$Z^{ab}_{ij}
\times Z^{cd}_{kl}$}{Z\^{}ab\_ij\xspace\texttimes
\xspace Z\^{}cd\_kl} single correlator bootstrap}
In Fig.\ \ref{ZZ_100_2} we display the bound for the maximum allowed
scaling dimension for the first operator in the $SY$ representation as a
function of $\Delta_Z$. For this plot we use the sum rules from Appendix~\ref{AppB}. There seems to be a kink roughly around the position of the
second $SY$ operator and first $Z$ operator of the corresponding
$\MNmath_{{100,2}}$ theory studied in \cite{Henriksson:2021lwn}. Beyond
this similarity at the level of certain scaling dimensions we do not know
if this kink is due to the same theory or some other theory. This is in
part due to the fact that \cite[Fig.\ 4]{Henriksson:2021lwn} converges very slowly with $\Lambda$.

\begin{figure}[H]
  \centering
  \includegraphics[scale=1]{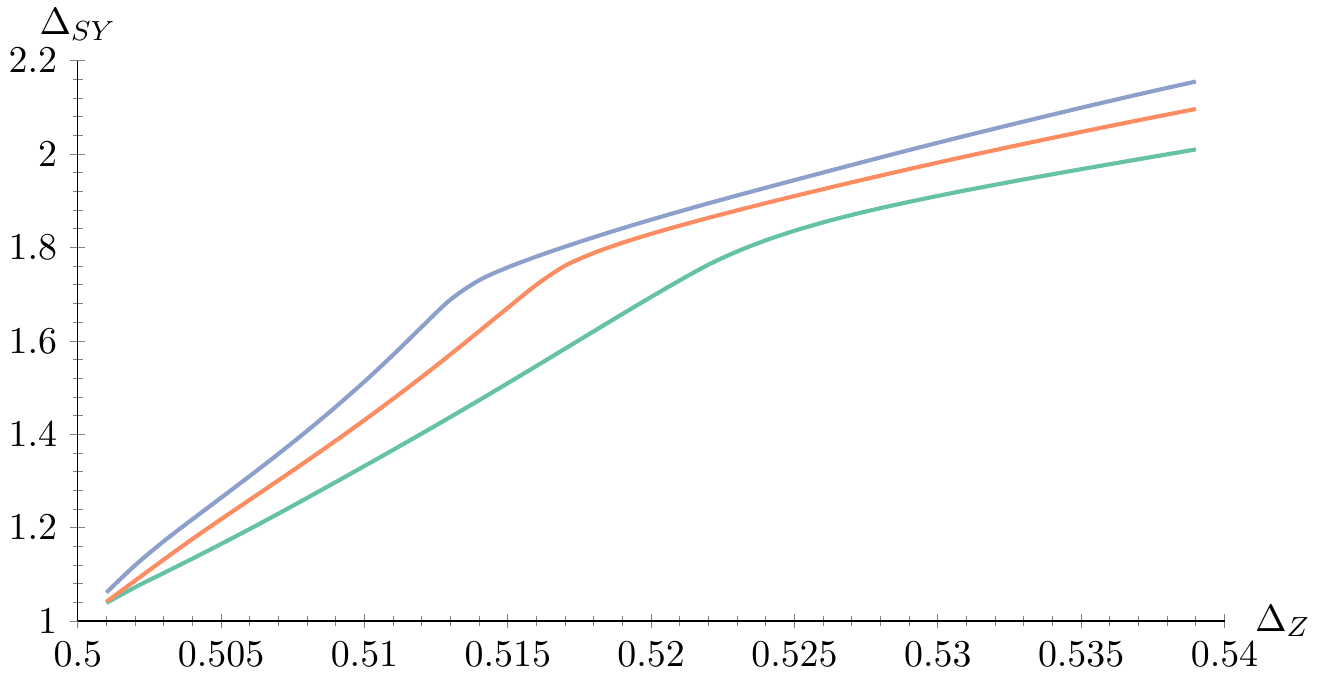}
  \caption{Single correlator \MN exclusion bound using the $n \geq 4$
  $\langle ZZZZ \rangle$ crossing equation. We display the behavior for
  $n=4$ fixed and for increasing $m$ ($m=10$ green, $m=100$ red and
  $m=1000$ blue).  The line corresponds to the maximum allowed scaling
  dimension of the first $SY$ (remember $SY=Y$) operator as a function of
  the scaling dimension of the order parameter operator. To obtain this
  figure with \texttt{qboot} we used $\Lambda = 15$, $\ell=\{0,\ldots, 50,
  55, 56, 59, 60, 64, 65, 69, 70, 74, 75, 79, 80, 84, 85, 89, 90\}$ and
  $\nu_{\max}=25$.}
  \label{ZZ_Large_M}
\end{figure}

\begin{figure}[H]
  \centering
  \includegraphics[scale=1]{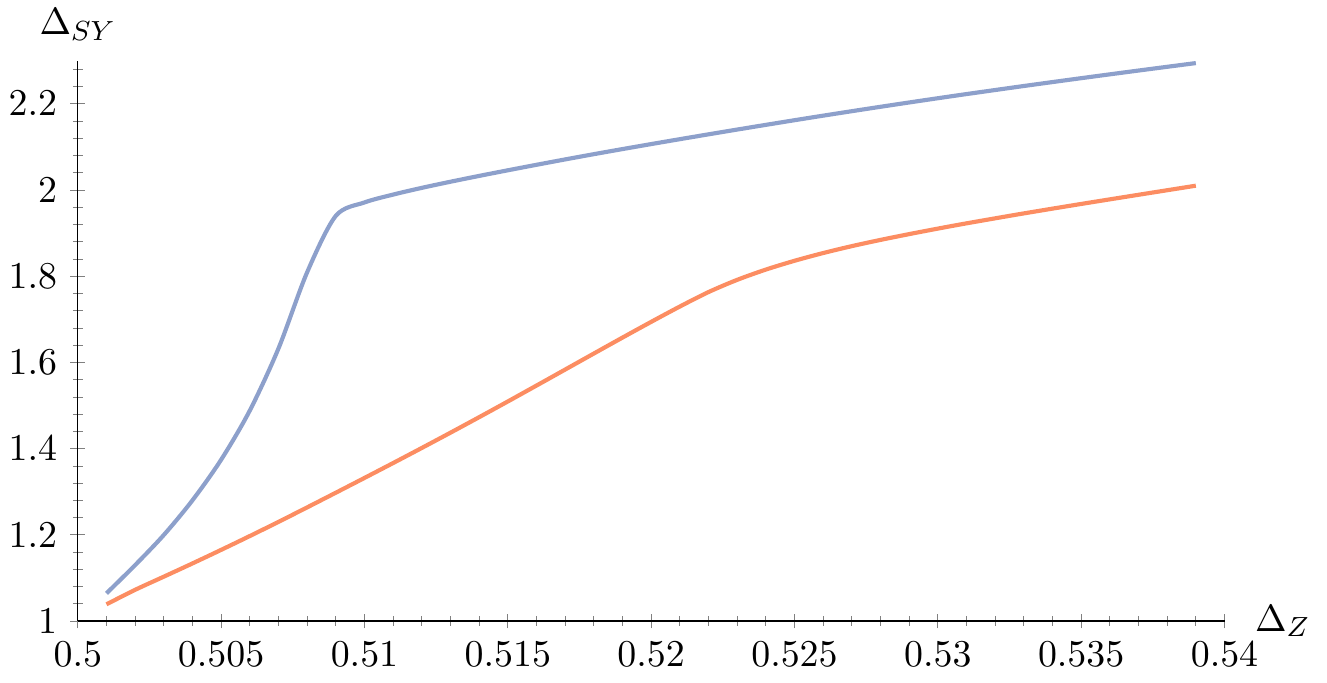}
  \caption{Single correlator \MN exclusion bound using the $n \geq 4$
  $\langle ZZZZ \rangle$ crossing equation. We display the behavior for
  $m=10$ fixed and for $n=4$ (red) and $n=10$ (blue). The line corresponds
  to the maximum allowed scaling dimension of the first $SY$ (remember
  $SY=Y$) operator as a function of the scaling dimension of the order
  parameter operator. To obtain this figure with \texttt{qboot} we used
  $\Lambda = 15$, $\ell=\{0, \ldots, 50, 55, 56, 59, 60, 64, 65, 69, 70,
  74, 75, 79, 80, 84, 85, 89, 90\}$ and $\nu_{\max}=25$.}
  \label{ZZ_Large_N}
\end{figure}

\subsec{The \texorpdfstring{$n\geq 4$}{n>= 4}
\texorpdfstring{$Z^{ab}_{ij} \times Z^{cd}_{kl}$}{Z\^{}ab\_ij\xspace
\texttimes\xspace Z\^{}cd\_kl} single correlator bootstrap}
With the expressions for the sum rules of a four-point function of
$Z^{ab}_{ij}$ operators worked out for generic $m \geq 2$ and $n \geq 4$,
it is interesting to study the behavior of the bounds in various parameter
limits. To this end, in Fig.\ \ref{ZZ_Large_M} we probe the large $m$
limit, whereas in Fig.\ \ref{ZZ_Large_N} we probe the large $n$ limit. We
find that the most interesting exclusion bound is the one in the $SY$
sector, which displays a sharp kink for all values of $m$ and $n$ tested.
In the large $n$ limit the kink converges to the point $(\Delta_{SY}=2$,
$\Delta_\phi=1$) which hints at a ``standard''\footnote{For the
``non-standard'' large $n$ description see the discussion in section
\ref{phi-X}} large $n$ Hubbard--Stratonovich description. On the other
hand, it is not clear what the theory converges to in the large $m$ limit.

\begin{figure}[H]
  \centering
  \includegraphics[scale=1]{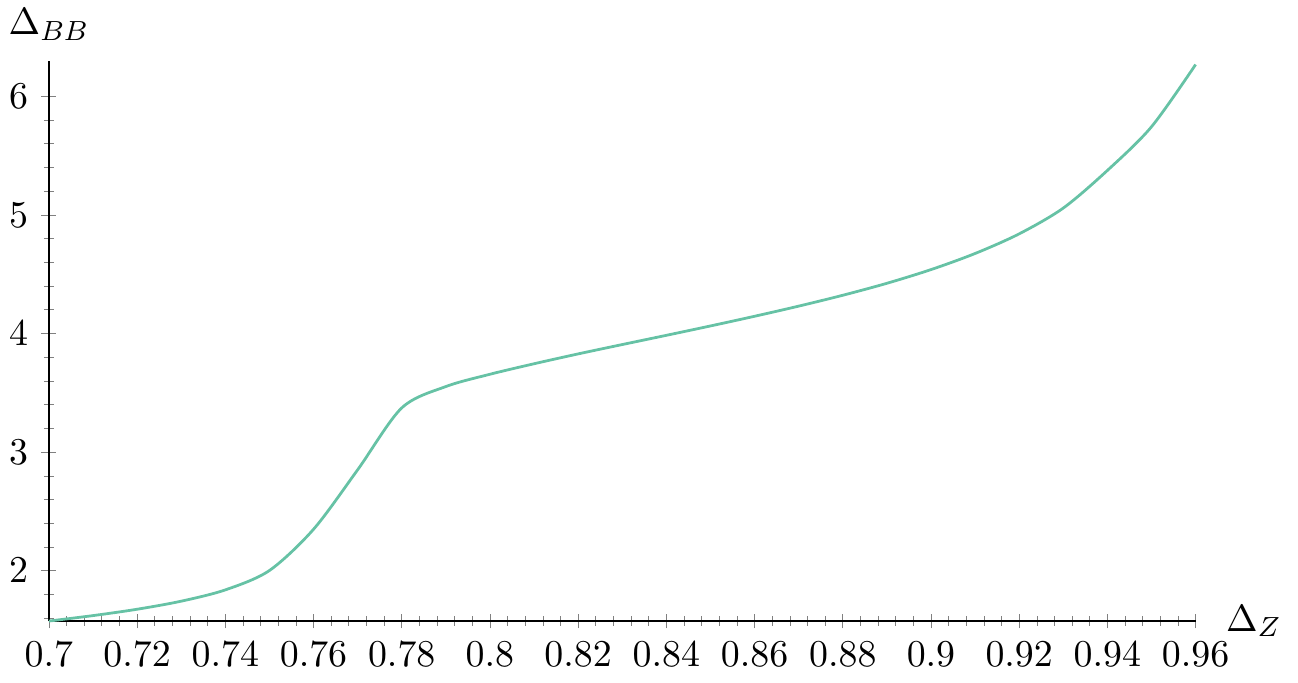}
  \caption{Single correlator \MN exclusion bound using the $n = 10$ and $m =2$
  $\langle ZZZZ \rangle$ crossing equation. The line corresponds
  to the maximum allowed scaling dimension of the first $BB$ operator in the $Z \times Z $ OPE. To obtain this figure with \texttt{qboot} we used
  $\Lambda = 15$, $\ell=\{0, \ldots, 50, 55, 56, 59, 60, 64, 65, 69, 70,
  74, 75, 79, 80, 84, 85, 89, 90\}$ and $\nu_{\max}=25$.}
  \label{ZZ_BB}
\end{figure}

Another interesting thing to look at is the exclusion bound for the first
operator transforming in the $BB$ representation. Representations that are
antisymmetric in two pairs of indices tend to require higher powers of the
field or derivatives in order to be written down and not vanish
identically;\footnote{We thank Alessandro Vichi for pointing us in this
direction.} see related discussions in \cite{He:2021xvg} and
\cite{Manenti:2021elk}. We note that this is true for theories where the
fields may be written as polynomials of other fields, i.e.\ which have a
weakly coupled description. However, we do expect intuitively that some
qualitative features may carry over to the strong coupling limit. This is
indeed what we observe in Fig.\ \ref{ZZ_BB} where we find a kink that has (very
roughly) $\Delta_{BB} \sim 2\Delta_{Z}+2 \sim 4\Delta_Z $.

Let us note that we have checked that our bounds do not change if we assume
that the external $Z$ operator appears in its OPE with itself. We checked
this by looking at the $\Delta_{SZ}\sim \Delta_{Z}$ exclusion bound, for
e.g.\ $m=5$ and $n=5$ and did not see any difference in our bounds (up to a
vertical precision of $10^{-6}$ that we checked). More explicitly, we checked that if $Z$ is exchanged in the $Z \times Z$ OPE, then the exclusion bound on the second exchanged operator in this representation is identical to the exclusion bound of the first exchanged operator in the $Z$ irrep if one assumes that the external operator is not exchanged. Additionally, we checked that for e.g. $m=1000$ and $n=4$ the corresponding $\Delta_{SY}$ exclusion bound in Fig.\ \ref{ZZ_Large_M} remains unchanged even after adding the assumption $\Delta_{SZ} \geq 1.0$. The question of whether
or not $Z$ should appear in its OPE with itself is important, since if it
does not, that would exclude a cubic (in powers of the order parameter
field) term in a possible Hamiltonian/Lagrangian description of the theory.
This is because in this case the $Z$ field would transform under an
additional $\mathbb{Z}_2$ symmetry.

\newsec{Summary and conclusion}[conc]
We have studied kinks and islands that arise when three-dimensional CFTs
with $O(m)^n\rtimes S_n$ global symmetry are analyzed with the numerical
bootstrap both ``close'' and ``far'' from the unitarity bound. With regards
to the analysis close to the unitarity bound, we found that in the large
number of copies limit the kinks converge to the expected position from
perturbative calculations.  However, when we examine theories with a small
number of copies we do not agree with the $\varepsilon$ expansion
predictions. This has two interpretations. The first interpretation is that
the theories we find are those of the $\varepsilon$ expansion, but for some
reason the actual perturbative predictions are very inaccurate. The second
is that the kinks are due to another theory that converges to the same
point when the number of copies is taken to be large. Note that this is
precisely the same qualitative behavior observed in the case of hypercubic
theories.

In addition to the above, we have worked out the tensor structures for
problems including groups of the form $G^n\rtimes S_n$ with $G$ and $n$
arbitrary. To this end, as an application of these results we bootstrapped
the four-point function of symmetric tensors $Z^{ab}_{ij}$ of $O(m)^n
\rtimes S_n$. Although we found various interesting features, we left a
more detailed analysis of these theories for future work.

There are various future directions that stem from the present work. One is
to bootstrap a mixed system of correlators involving $\phi^a_i$ and
$Z^{bc}_{jk}$, this would give us a better handle on theories like the ones
in Figs.\ \ref{MN_20_2_island} and \ref{ZZ_100_2}. Also, having worked out
the tensor structures for $G^n\rtimes S_n$ it would be interesting to
bootstrap the theories consisting of $n$ copies of the $m$ state Potts
models and see if we can find any interesting features; see e.g.
\cite{Dotsenko:1998gyp}.

\ack{We thank Mocho Go for assistance regarding the implementation of OPE
coefficient relations and definitions in \texttt{qboot}. We are grateful to
Alessandro Vichi for useful discussions and suggestions and to Johan Henriksson for reading through the manuscript and providing useful comments. Additionally, we thank two anonymous referees for their careful reading and constructive comments that helped improve this manuscript.
Numerical computations in this paper were run on the Crete Center for
Quantum Complexity and Nanotechnology. This research used resources
provided by the Los Alamos National Laboratory Institutional Computing
Program, which is supported by the U.S.\ Department of Energy National
Nuclear Security Administration under Contract No.\ 89233218CNA000001. This
research used resources of the National Energy Research Scientific
Computing Center (NERSC), a U.S.\ Department of Energy Office of Science
User Facility operated under Contract No.\ DE-AC02-05CH11231.

The research work of SRK was supported by the Hellenic Foundation for
Research and Innovation (HFRI) under the HFRI PhD Fellowship grant
(Fellowship Number: 1026). The research work of SRK also received funding from the European Research Council (ERC) under the European Union’s Horizon 2020 research and innovation programme (grant agreement no. 758903).

Research presented in this article was supported by the Laboratory Directed
Research and Development program of Los Alamos National Laboratory under
project number 20180709PRD1. AS is funded by the Royal Society under the
grant ``Advancing the Conformal Bootstrap Program in Three and Four
Dimensions''.

\vspace{-12pt}
\begin{figure}[H]
  \flushright
  \includegraphics[scale=0.45]{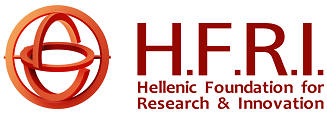}
\end{figure}
}
\vspace{-1.5cm}

\begin{appendices}

\section{\texorpdfstring{$\boldsymbol{\phi\text{-}X}$}{phi-X} system sum
rules}\label{appA}
In this appendix we collect the sum rules that result from each crossing
equation in the system of $\phi$-$X$ mixed correlator equations. We define
$F_{\pm , \Delta, l}^{ij,kl}$ as in \cite{Kos:2015mba} and we also use
$F^{\pm}$ and $F_{\pm}$ interchangeably.
\subsection{\texorpdfstring{$\langle \phi \phi \phi \phi
\rangle$}{\textlangle phi phi phi phi
\textrangle} crossing equation}
The single correlator sum rules have already appeared in
\cite{Stergiou:2019dcv} and \cite{Henriksson:2021lwn}. We quote them below
for completeness:
\begin{align}
&\sum_{S^+}c_O^2\begin{pmatrix}
  0\\
  F^-_{\Delta,\lsp\ell}\\
  0\\
  0\\
  F^+_{\Delta,\lsp\ell}\\
  0
\end{pmatrix}+
\sum_{ X^+}c_O^2\begin{pmatrix}
  0\\
  -F^-_{\Delta,\lsp\ell}\\
  n \lsp F^-_{\Delta,\lsp\ell}\\
  0\\
  -\lsp F^+_{\Delta,\lsp\ell}\\
  n \lsp F^+_{\Delta,\lsp\ell}
\end{pmatrix}+
\sum_{Y^+}c_O^2\begin{pmatrix}
  0\\
  0\\
  \frac{m-2}{2}\lsp F^-_{\Delta,\lsp\ell}\\
  m \lsp F^-_{\Delta,\lsp\ell}\\
  0\\
  \frac{m-2}{2}\lsp F^+_{\Delta,\lsp\ell}
\end{pmatrix}
+\sum_{Z^+}c_O^2\begin{pmatrix}
  F^-_{\Delta,\lsp\ell}\\
  \frac{1}{2} \lsp F^-_{\Delta,\lsp\ell}\\
  -\frac{1}{2} \lsp F^-_{\Delta,\lsp\ell}\\
  -F^-_{\Delta,\lsp\ell}\\
  -\frac{1}{2} \lsp F^+_{\Delta,\lsp\ell}\\
  +\frac{1}{2} \lsp F^+_{\Delta,\lsp\ell}
\end{pmatrix}\nonumber\\
&\hspace{6.5cm}+
\sum_{A^-}c_O^2\begin{pmatrix}
  0\\
  0\\
  -\frac{1}{2}\lsp F^-_{\Delta,\lsp\ell}\\
  F^-_{\Delta,\lsp\ell}\\
  0\\
  \frac{1}{2}\lsp F^+_{\Delta,\lsp\ell}
\end{pmatrix}+
\sum_{B^-}c_O^2\begin{pmatrix}
  F^-_{\Delta,\lsp\ell}\\
  -\frac{1}{2}\lsp F^-_{\Delta,\lsp\ell}\\
  \frac{1}{2}\lsp F^-_{\Delta,\lsp\ell}\\
  -F^-_{\Delta,\lsp\ell}\\
  \frac{1}{2}\lsp F^+_{\Delta,\lsp\ell}\\
  -\frac{1}{2}\lsp F^+_{\Delta,\lsp\ell}
\end{pmatrix}=
\begin{pmatrix}
  0\\
  0\\
  0\\
  0\\
  0\\
  0
\end{pmatrix}.
\label{mncrossing}
\end{align}

\subsection{\texorpdfstring{$\langle \phi X\xspace\phi X \rangle$}{\textlangle
phi Xphi X \textrangle} crossing equation}
Using the projectors \eqref{aeq7} and \eqref{aeq6} we find the sum rules
\begin{equation}
\sum_O \lambda_{\phi X O_{\bar{A}}}^2F_{\mp,\Delta ,l}^{ \phi X , \phi X}
=0\,,
\end{equation}
and
\begin{equation}
  \sum_O \lambda_{\phi X O_{\phi}}^2F_{\mp,\Delta ,l}^{\phi X , \phi X}
  =0\,.
\end{equation}

\subsection{\texorpdfstring{$\langle \phi  \phi XX \rangle$}{\textlangle
phiphi XX \textrangle} crossing equation}
For the crossing equation from the four-point function $ \langle \phi  \phi
X X \rangle$ we also need the projectors corresponding to the decomposition
of $ \langle \phi^a_i  \phi^b_j X^{cd} X^{ef} \rangle$ onto irreps. To find
these we need to know the irreps the $\phi \times \phi$ and $X \times X$
OPEs have in common. These are the $X$ and $S$ irreps. The corresponding
projectors turn out to be \cite{Kousvos:2018rhl, Kousvos:2021mpw}
\begin{equation}
P^S_{abcdef}=\delta_{ab}\Big(\delta_{cdef}-\frac{1}{n}\delta_{cd}\delta_{ef}
\Big)
\label{eq32}
\end{equation}
and
\begin{equation}
  P^X_{abcdef}=\delta_{abcdef}-\frac{1}{n}\Big(\delta_{ab}\delta_{cdef}+\delta_{cd}\delta_{abef}+\delta_{ef}\delta_{abcd}\Big)+\frac{2}{n^2}\delta_{ab}\delta_{cd}\delta_{ef}\,.
\label{eq33}
\end{equation}
We thus obtain the sum rules
\begin{equation}
\sum_O(\lambda_{\phi \phi O_{S+}}\lambda_{XX O_{S+}} F_{\mp,\Delta ,
l}^{\phi\phi, XX} \pm (-1)^l \frac{1}{n}\lambda_{\phi X O_{y}}^2
F_{\mp,\Delta ,l}^{X\phi, \phi X}\pm (-1)^l \frac{1}{n}\lambda_{\phi X O_{\bar{A}}}^2F_{\mp,\Delta ,l}^{ X\phi, \phi X} )=0
\end{equation}
and
\begin{equation}
\sum_O(\lambda_{\phi \phi O_{X+}}\lambda_{XX O_{X+}} F_{\mp,\Delta ,
l}^{\phi\phi, XX} \pm (-1)^l \lambda_{\phi X O_{y}}^2 F_{\mp,\Delta ,l}^{ X
\phi, \phi X}\mp (-1)^l \lambda_{\phi X O_{\bar{A}}}^2F_{\mp,\Delta ,l}^{ X
\phi, \phi X} )=0\,.
\end{equation}

\subsection{\texorpdfstring{$\langle  X XX X \rangle$}{\textlangle
XXXX\textrangle} crossing equation}
For the $\langle XXXX \rangle$ crossing equation the sum rules have been
already computed in the literature, albeit in a slightly different context,
see e.g.\ \cite{Rong:2017cow} and \cite{Stergiou:2018gjj}. We quote
\begin{align}
&\sum_{S^+}c_O^2\begin{pmatrix}
  0\\
  F^-_{\Delta,\lsp\ell}\\
  F^+_{\Delta,\lsp\ell}\\
  0
\end{pmatrix}+
\sum_{ X^+}c_O^2\begin{pmatrix}
  0\\
  0\\
  -\frac{4}{n+1}\lsp F^+_{\Delta,\lsp\ell}\\
   \lsp F^-_{\Delta,\lsp\ell}
\end{pmatrix}+
\sum_{E^+}c_O^2\begin{pmatrix}
  \lsp F^-_{\Delta,\lsp\ell}\\
  \frac{2(n-1)}{n} \lsp F^-_{\Delta,\lsp\ell}\\
  -\frac{(n+1)(n-2)}{n(n-1)}\lsp F^+_{\Delta,\lsp\ell}\\
  -\frac{n+1}{2(n-1)}\lsp F^-_{\Delta,\lsp\ell}
\end{pmatrix}
+\sum_{\bar{S}^-}c_O^2\begin{pmatrix}
  F^-_{\Delta,\lsp\ell}\\
  0\\
  F^+_{\Delta,\lsp\ell}\\
  0
\end{pmatrix}=
\begin{pmatrix}
  0\\
  0\\
  0\\
  0
\end{pmatrix}.
\label{tetrahedralcrossing}
\end{align}

\subsubsection{The \texorpdfstring{$n=3$}{n=3} case}
For $n=3$ the sum rules become
\begin{align}
\sum_{S^+}c_O^2\begin{pmatrix}
  0\\
  F^-_{\Delta,\lsp\ell}\\
  F^+_{\Delta,\lsp\ell}
\end{pmatrix}+
\sum_{ X^+}c_O^2\begin{pmatrix}
  F^-_{\Delta,\lsp\ell}\\
  0\\
   -2\lsp F^+_{\Delta,\lsp\ell}
\end{pmatrix}+
\sum_{\bar{S}^-}c_O^2\begin{pmatrix}
  F^-_{\Delta,\lsp\ell}\\
  - \lsp F^-_{\Delta,\lsp\ell}\\
  F^+_{\Delta,\lsp\ell}
\end{pmatrix}\nonumber =
\begin{pmatrix}
  0\\
  0\\
  0
\end{pmatrix},
\label{O2crossing}
\end{align}
which are equivalent to those of $O(2)$.

\subsection{\texorpdfstring{$n=2$}{n=2} sum rules}
With the exception of the single correlator sum rules, all the above sum rules are drastically simplified when $n=2$, this is because the irrep $X$ becomes one dimensional then.\footnote{Hence, all OPEs involving $X$ only exchange one irrep.} We thus have

\begin{equation}
  \sum_{S^+}\lambda_{XX O_S}^2 F^{- XX,XX}_{\Delta , l} =0\,,
\label{ceq5}
\end{equation}

\begin{equation}
  \sum_{\phi^\pm}\lambda_{O XO_\phi}^2 F^{- OX,OX}_{\Delta , l} =0\,,
\label{ceq6}
\end{equation}

\begin{equation}
\sum_{S^+}\lambda_{O O O_S} \lambda_{ X X O_S}F^{\mp OO,XX}_{\Delta , l}\pm
\sum_{\phi^\pm}(-1)^l\lambda_{O XO_\phi}^2 F^{\mp XO,OX}_{\Delta , l} =0\,,
\label{ceq7}
\end{equation}
in addition to the single correlator sum rules which remain as defined earlier.

\section{\texorpdfstring{$\boldsymbol{Z\times Z}$}{Z\texttimes Z} single correlator sum rules}
\label{AppB}
The $n=2$ $Z \times Z$ bootstrap the sum rules are simplified drastically by the observation that $Z^{ab}_{ij}$ and $Z^{cd}_{kl}$ necessarily have ($a=c$ and $b=d$) or ($a=d$ and $b=c$). Thus if we assume that the field $Z^{ab}_{ij}$ has, lets say, $a=1$ then by definition $b=2$. Hence, the projectors are
\begin{align}
\begin{split}
{P^S}_{ijklmnop}&= P^S_{ikmo}P^S_{jlnp}\,,\\
{P^{SY}}_{ijklmnop}&=P^S_{ikmo}P^Y_{jlnp}+P^Y_{ikmo}P^S_{jlnp}\,,\\
{P^{SA}}_{ijklmnop}&=P^S_{ikmo}P^A_{jlnp}+P^A_{ikmo}P^S_{jlnp}\,,\\
{P^{YA}}_{ijklmnop}&=P^Y_{ikmo}P^A_{jlnp}+P^A_{ikmo}P^Y_{jlnp}\,,\\
{P^{YY}}_{ijklmnop}&=P^Y_{ikmo}P^Y_{jlnp}\,,\\
{P^{AA}}_{ijklmnop}&=P^A_{ikmo}P^A_{jlnp}\,,
\end{split}
\label{ZZprojectors}
\end{align}
i.e.\ the products of projectors of $O(m)$. By $S$ we denote the scalar
singlet, by $A$ the antisymmetric and by $Y$ the rank-two traceless symmetric
irreps of $O(m)$. We obtain the following sum rules
\begin{align}
&\sum_{S^+}c_O^2\begin{pmatrix}
  0\\
  \lsp F^-_{\Delta,\lsp\ell}\\
  0\\
  0\\
  0\\
  \lsp F^+_{\Delta,\lsp\ell}
\end{pmatrix}+
\sum_{ SY^+}c_O^2\begin{pmatrix}
  0\\
  -2\lsp F^-_{\Delta,\lsp\ell}\\
  \lsp F^-_{\Delta,\lsp\ell}\\
  2\lsp F^-_{\Delta,\lsp\ell}\\
  -2\lsp F^+_{\Delta,\lsp\ell}\\
  \frac{m-4}{2}\lsp F^+_{\Delta,\lsp\ell}
\end{pmatrix}+
\sum_{YY^+}c_O^2\begin{pmatrix}
  \lsp F^-_{\Delta,\lsp\ell}\\
  \lsp F^-_{\Delta,\lsp\ell}\\
  -\lsp F^-_{\Delta,\lsp\ell}\\
  (m-2)\lsp F^-_{\Delta,\lsp\ell}\\
  (m+2)\lsp F^+_{\Delta,\lsp\ell}\\
  \frac{2-m-m^2}{2}\lsp F^+_{\Delta,\lsp\ell}
\end{pmatrix}
+\sum_{AA^+}c_O^2\begin{pmatrix}
  \lsp F^-_{\Delta,\lsp\ell}\\
  0\\
  0\\
  0\\
  -m\lsp F^+_{\Delta,\lsp\ell}\\
  0
\end{pmatrix}\nonumber\\
&\hspace{6.5cm}+
\sum_{SA^-}c_O^2\begin{pmatrix}
  0\\
  0\\
  -\lsp F^-_{\Delta,\lsp\ell}\\
  0\\
  -2\lsp F^+_{\Delta,\lsp\ell}\\
  \frac{m}{2}\lsp F^+_{\Delta,\lsp\ell}
\end{pmatrix}+
\sum_{YA^-}c_O^2\begin{pmatrix}
  2\lsp F^-_{\Delta,\lsp\ell}\\
  -m^2\lsp F^-_{\Delta,\lsp\ell}\\
  \lsp F^-_{\Delta,\lsp\ell}\\
  m\lsp F^-_{\Delta,\lsp\ell}\\
  2\lsp F^+_{\Delta,\lsp\ell}\\
  \frac{m^2-m}{2}\lsp F^+_{\Delta,\lsp\ell}
\end{pmatrix}=
\begin{pmatrix}
  0\\
  0\\
  0\\
  0\\
  0\\
  0
\end{pmatrix}.
\label{zzcrossing}
\end{align}

%
%
%
\section{\texorpdfstring{$\boldsymbol{\langle ZZZZ \rangle}$}{\textlangle
  ZZZZ\textrangle} projectors for \texorpdfstring{$\boldsymbol{n \geq
  4}$}{n>= 4} and \texorpdfstring{$\boldsymbol{G}$}{G} arbitrary}
\label{zz_group_theory}
The projectors that correspond to the $\langle Z^{ab}_{ij} Z^{cd}_{kl}
Z^{ef}_{mn} Z^{gh}_{op} \rangle$ correlator are rather extended 16 index
objects. In order to simplify their presentation we will introduce
``Blocks'', not to be confused with conformal blocks. With the use of
Blocks the projectors can be presented in somewhat compact form. Also, we
will make no distinction between upper and lower indices. It will be
implicitly assumed that indices $a$-$h$ label the copy of $G$, and the
indices $i$-$p$ are $G$ indices. This setup is not the most general
possible with respect to $G$ indices, but we use it for simplicity of
demonstration. We hope that the generalization to arbitrary indices of $G$
will become obvious from our presentation. There are three groups of
representations that appear in the $Z^{ab}_{ij} \times Z^{cd}_{kl}$ OPE.
These are:\\
\\
Group I: Representations with $a$ = $c$ or $d$ OR $b$ = $c$ or $d$.\\
\\
Group II: Representations with $a$ = $c$ or $d$ AND $b$ = $d$ or $c$.\\
\\
Group III: Representations with $a$, $b$, $c$ and $d$ all different.\\
\\
To reiterate, we have Group I: two pairs of indices equal, Group II: one pair of indices equal, Group III: no pairs of indices equal. We also remind the reader that the operators $Z^{ab}_{ij}$ have $a \neq b$.

\subsection{Group I representations}
Let us start by recalling the projectors of the $G$ irreps (we take
$G=O(m)$ in order to be explicit, but the generalization to any $G$ is
trivial). We have $P_{g_1}^{ijkl}=\frac{1}{m}\delta^{ij}\delta^{kl}$,
$P_{g_2}^{ijkl}=\frac{1}{2}(\delta^{ik}\delta^{jl}-\delta^{il}\delta^{jk})$
and $P_{g_3}^{ijkl} =
\frac{1}{2}(\delta^{ik}\delta^{jl}+\delta^{il}\delta^{jk}-\frac{2}{m}\delta^{ij}\delta^{kl})$.
Next we define the following useful tensors:
\begin{align}
\begin{split}
  {R_1}_{abcdefgh}&=(\delta_{ac}\delta_{bd}-\delta_{acbd})(\delta_{eg}\delta_{fh}-\delta_{egfh})\,,\\
  {R_2}_{abcdefgh}&=\delta_{aceg}\delta_{bd}\delta_{fh}-\delta_{abcdeg}\delta_{fh}-\delta_{afcheg}\delta_{bd}+\delta_{abcdefgh}\,,\\
  {R_3}_{abcdefgh}&=\delta_{aceg}\delta_{bdfh}-\delta_{abcdefgh}\,.
\end{split}
\label{threetensors}
\end{align}
With these tensors in hand we can now define the Blocks. We denote these
with ``$B$'' for short. They are
\begin{align}
\begin{split}
{B_S}^{abcdefgh}_{ijklmnop} &= P_{g_1}^{i k m o} P_{g_1}^{j l n
  p}  {R_1}_{abcdefgh}\,,\\
{B_{XS}}^{abcdefgh}_{ijklmnop} &= P_{g_1}^{i k m o} P_{g_1}^{j l n
  p} \Big({R_2}_{a b c d e f gh} - \frac{1}{n} {R_1}_{abcdefgh}\Big)\,,\\
{B_{XY}}^{abcdefgh}_{ijklmnop} &= P_{g_1}^{i k m o} P_{g_3}^{j l n
  p} \Big({R_3}_{a b c d e f gh} - \frac{1}{n-1} {R_2}_{b a d c f e h
  g}\Big)\,,\\
{B_{XA}}^{abcdefgh}_{ijklmnop} &= P_{g_1}^{i k m o} P_{g_2}^{j l n
  p} \Big({R_3}_{a b c d e f gh} - \frac{1}{n-1} {R_2}_{b a d c f e h
  g}\Big)\,,\\
{B_{SY}} ^{abcdefgh}_{ijklmnop}&= P_{g_1}^{i k m o} P_{g_3}^{j l n p} {R_2}_{b a d c f e h g}\,,\\
{B_{SA}} ^{abcdefgh}_{ijklmnop}&= P_{g_1}^{i k m o} P_{g_2}^{j l n p} {R_2}_{b a d c f e h g}\,,\\
{B_{YY}}^{abcdefgh}_{ijklmnop} &= P_{g_3}^{i k m o} P_{g_3}^{j l n p} {R_3}_{b a d c f e h g}\,,\\
{B_{AA}} ^{abcdefgh}_{ijklmnop}&= P_{g_2}^{i k m o} P_{g_2}^{j l n p} {R_3}_{b a d c f e h g}\,,\\
{B_{Y\!A}} ^{abcdefgh}_{ijklmnop}&= P_{g_3}^{i k m o} P_{g_2}^{j l n p} {R_3}_{b a d c f e h g}\,,\\
{B_{\bar{X}}} ^{abcdefgh}_{ijklmnop}&= P_{g_1}^{i k m o} P_{g_1}^{j l n
p}\Big({R_3}_{b a d c f e h  g} -\frac{1}{n-2} ({R_2}_{a b c defg h} +
{R_2}_{b a d c f e h g}) \\
&\hspace{6cm}+ \frac{1}{(n-1)(n-2)} {R_1}_{a b c d e f g h}\Big)\,.
\end{split}
\label{blocks}
\end{align}
To finally obtain the projectors from the Blocks we must perform
symmetrizations, which are the same for all Blocks, hence we write $B_g$,
where $g$ is one of the irreps considered above.  These symmetrizations are
\begin{align}
\begin{split}
  {{{B_g}'} \lsp}^{abcdefgh}_{ijklmnop} &= {{{B_g}} \lsp}^{a b c d e f g
h}_{ijklmnop} + {{{B_g}} \lsp}^{a b c d e f h g}_{i j k l m n p o} \,,\\
{{{B_g}''} \lsp}^{abcdefgh}_{ijklmnop} &= {{{B_g}'}
\lsp}^{abcdefgh}_{ijklmnop} + {{{B_g}'} \lsp}^{a b c d f e g h }_{i j k
l n m o p}\,,\\
{{B_g}'''\lsp}^{abcdefgh}_{ijklmnop} &=
{{B_g}'' \lsp}^{abcdefgh}_{ijklmnop}+
{{B_g}''\lsp}^{a b d c e f, g h }_{i j l k m n o p}\,,\\
{{P_g}\lsp}^{abcdefgh}_{ijklmnop} &= {{{B_g}'''
}\lsp}^{abcdefgh}_{ijklmnop} +{{B_g}'''}{\lsp}^{b a c
d e f g h }_{j i k l m n o p}\,,
\end{split}
\label{symmetrizations}
\end{align}
where the last line in \eqref{symmetrizations} corresponds to the final expression for the projector in irrep $g$. Note that the above symmetrizations are necessary in order to take into account the symmetry $Z^{ab}_{ij}=Z^{ba}_{ji}$. To apply these equations to a different group $G$ one simply needs to replace the expressions $P_{g_1}$, $P_{g_2}$ and $P_{g_3}$ with those of their group of choice.

\subsection{Group II representations}
For Group II representations the steps are very similar with the ones described for Group I. We must simply define some new tensors. These are
\begin{equation}
\begin{aligned}
{RR_1}_{a b c d e f g  h} &= \delta_{a c eg} (\delta_{bf} \delta_{dh}   -\delta_{bfdh}) - (\delta_{acegbf}\delta_{dh} -\delta_{acegbfdh}) - (\delta_{acegdh}\delta_{bf} - \delta_{acegbfdh})\\
{RR_2}_{a b c d e f g  h}   &=\delta_{ac}\delta_{eg} ( \delta_{bf}\delta_{dh} - \delta_{bfdh})
 - \delta_{eg} (\delta_{acbf}\delta_{dh} +\delta_{acdh}\delta_{bf} -
 2\delta_{acbfdh})\,,\\
 &\quad-\delta_{ac} (\delta_{egbf}\delta_{dh} +\delta_{egdh}\delta_{bf} -
 2\delta_{egbfdh})
 + (\delta_{acbfeg}\delta_{dh} +\delta_{acdheg}\delta_{bf} -
    2\delta_{acegbfdh})\\
    &\quad+ (\delta_{acbf}\delta_{egdh} +\delta_{acdh}\delta_{bfeg} -
    2\delta_{acegbfdh})\,,
\end{aligned}
\end{equation}
whereas the Blocks for Group II representations are
\begin{equation}
\begin{aligned}
{B_{XB}}^{abcdefgh}_{ijklmnop}&=({RR_1}_{a b c d e f g h}-\frac{1}{n-2}{RR_2}_{a b c d e f g
h})P_{g_1}^{i k m o}\delta_{jn}\delta_{lp}\,,\\
&\quad-({RR_1}_{a b c d e h g f}-\frac{1}{n-2}{RR_2}_{a b c d e h g f})
P_{g_1}^{i k m o}\delta_{jp}\delta_{ln}\,,\\
{B_{XZ}}^{abcdefgh}_{ijklmnop}&=({RR_1}_{a b c d e f g h}-\frac{1}{n-2}{RR_2}_{a b c d e f g
h})P_{g_1}^{i k m o}\delta_{jn}\delta_{lp}\,,\\
&\quad+({RR_1}_{a b c d e h g f}-\frac{1}{n-2}{RR_2}_{a b c d e h g f})
P_{g_1}^{i k m o}\delta_{jp}\delta_{ln}\,,\\
{B_{SB}}^{abcdefgh}_{ijklmnop}&={RR_2}_{a b c d e f g h} P_{g_1}^{i k m o}\delta_{jn}\delta_{lp}
-  {RR_2}_{a b c d e h g f} P_{g_1}^{i k m o}\delta_{jp}\delta_{ln}\,,\\
{B_{SZ}}^{abcdefgh}_{ijklmnop}&={RR_2}_{a b c d e f g h} P_{g_1}^{i k m o}\delta_{jn}\delta_{lp}
+  {RR_2}_{a b c d e h g f} P_{g_1}^{i k m o}\delta_{jp}\delta_{ln}\,,\\
{B_{YB}}^{abcdefgh}_{ijklmnop}&= {RR_1}_{a b c d e f g h} P_{g_3}^{i k m o}\delta_{jn}\delta_{lp}
-  {RR_1}_{a b c d e h g f} P_{g_3}^{i k m o}\delta_{jp}\delta_{ln}\,,\\
{B_{YZ}}^{abcdefgh}_{ijklmnop}&= {RR_1}_{a b c d e f g h} P_{g_3}^{i k m o}\delta_{jn}\delta_{lp}
+  {RR_1}_{a b c d e h g f} P_{g_3}^{i k m o}\delta_{jp}\delta_{ln}\,,\\
{B_{AB}}^{abcdefgh}_{ijklmnop}&= {RR_1}_{a b c d e f g h} P_{g_2}^{i k m o}\delta_{jn}\delta_{lp}
-  {RR_1}_{a b c d e h g f} P_{g_2}^{i k m o}\delta_{jp}\delta_{ln}\,,\\
{B_{AZ}}^{abcdefgh}_{ijklmnop}&= {RR_1}_{a b c d e f g h} P_{g_2}^{i k m o}\delta_{jn}\delta_{lp}
+  {RR_1}_{a b c d e h g f} P_{g_2}^{i k m o}\delta_{jp}\delta_{ln}\,.\\
\end{aligned}
\end{equation}
We may now apply \eqref{symmetrizations} and obtain the final expression for the projectors.

\subsection{Group III representations}
For the last group of representations we need to define one more tensor
structure:
\begin{equation}
\begin{aligned}
{R_4}^{abcdefgh}_{ijklmnop}&=(\delta_{ae}\delta_{bf}\delta_{cg}\delta_{dh}-
   \delta_{cg}\delta_{dh}\delta_{abef} -
   \delta_{bf}\delta_{dh}\delta_{aceg} -
   \delta_{bf}\delta_{cg}\delta_{adeh} -
   \delta_{ae}\delta_{dh}\delta_{bcfg} \\
   &\quad-\delta_{ae}\delta_{cg}\delta_{bdfh} -
   \delta_{ae}\delta_{bf}\delta_{cdgh} +  \delta_{aedh}\delta_{bcfg} +
   \delta_{aecg} \delta_{bdfh} + \delta_{aebf} \delta_{cdgh}\\
   &\quad+
   2 (\delta_{dh}\delta_{abcefg} +
      \delta_{cg}\delta_{abdefh} +
      \delta_{bf}\delta_{adcehg} +
      \delta_{ae}\delta_{dbchfg} ) -
    6  \delta_{aecgbfdh}) \delta_{im} \delta_{ko}  \delta_{jn}
    \delta_{lp}\,,
 \end{aligned}
\end{equation}
leading to the Blocks
\begin{equation}
\begin{aligned}
{B_{BB}\lsp} ^{abcdefgh}_{ijklmnop}&= {R_4}^{abcdefgh}_{ijklmnop} - {R_4}^{a
b c d g f e h}_{ i j k l o n m p} - {R_4}^{a b c d e h g f}_{ i j k l m p o
n} + {R_4}^{a b c d g h e f}_{ i j k l o p m n}\,,\\
{B_{BZ}\lsp} ^{abcdefgh}_{ijklmnop}&={R_4}^{abcdefgh}_{ijklmnop} - {R_4}^{a b
c d g f e h}_{ i j k l o n m p} + {R_4}^{a b c d e h g f}_{ i j k l m p o
n} - {R_4}^{a b c d g h e f}_{ i j k l o p m n}\,,\\
{B_{\TotS}\lsp} ^{abcdefgh}_{ijklmnop}& = {R_4}^{abcdefgh}_{ijklmnop} +
{R_4}^{a b c d e f h g }_{i j kl m n p o}\,.
\end{aligned}
\end{equation}
For the representations $BB$ and $BZ$ we may simply use
\eqref{symmetrizations} to obtain the projectors. On the other hand, $\TotS$ needs a slightly more elaborate formula, which is due to our explicit choice of Block. We have
\begin{equation}
\begin{aligned}
{{{B_{\TotS}}^\prime }\lsp}^{abcdefgh}_{ijklmnop} &= {B_{\TotS}\lsp}
^{abcdefgh}_{ijklmnop} + {B_{\TotS}\lsp} ^{a b c d e g f h}_{ i j k l m o n p}
+ {B_{\TotS}\lsp} ^{a b c d e h g f}_{i j k l m p o n}\,,\\
{{{P_{\TotS}} }\lsp}^{abcdefgh}_{ijklmnop} &= {{{B_{\TotS}}^\prime
}\lsp}^{abcdefgh}_{ijklmnop} +  {{{B_{\TotS}}^\prime }\lsp}^{a b c d f e g
h}_{ i j k l n m o p}  +  {{{B_{\TotS}}^\prime }\lsp}^{a b c d g f e h}_{ i
j k l o n m p}  +  {{{B_{\TotS}}^\prime }\lsp}^{a b c d h f g e}_{ i j k l p
n o m}\,.
\end{aligned}
\end{equation}

\subsection{Comments}
The above tensors can be derived if we know the symmetry properties of each
irrep (symmetric, antisymmetric, traceless, trace, \dots), by writing down a
tensor with the same symmetry and then contracting with a tensor that
enforces the constraints $a\neq b$, $c \neq d$, $e \neq f$ and $g \neq h$.
Such a tensor can be found in a straightforward way. First, define a
reduced Kronecker delta $\delta_{ab}^r = \delta_{ab} - \delta_{abr}$, which
is the same as the usual Kronecker delta, but equal to zero if the indices
are equal to a value $r$. Now we can define $P^{abcd}_{a^\prime b^\prime
c^\prime d^\prime } = \delta_{aa^\prime}^b \delta_{b b^\prime}
\delta_{cc^\prime}^d \delta_{d d^\prime} $ which is simplified as
$P^{abcd}_{a^\prime b^\prime c^\prime d^\prime } =
(\delta_{aa^\prime}\delta_{bb^\prime}-\delta_{aa^\prime
bb^\prime})(\delta_{cc^\prime}\delta_{dd^\prime}-\delta_{cc^\prime
dd^\prime})$ by plugging in the definition of the reduced Kronecker delta.
Now suppose that we had found a tensor $T^{a^\prime b^\prime c^\prime
d^\prime e^\prime f^\prime g^\prime h^\prime}$ with all the required
symmetry properties to describe some irrep. We could turn it into a
projector using the following equation:
\begin{equation}
T^{abcdefgh}= P^{abcd}_{a^\prime b^\prime c^\prime d^\prime }
P^{efgh}_{e^\prime f^\prime g^\prime h^\prime }T^{a^\prime b^\prime
c^\prime d^\prime e^\prime f^\prime g^\prime h^\prime}\,,
\end{equation}
where the left-hand side is our projector. One can also define generalized
reduced Kronecker deltas in order to impose more complicated constraints.
For example, let us consider a tensor with four indices which must take
different values and is totally symmetric (this is relevant for the $\TotS$
irrep). We may start with a tensor $T^{a^\prime b^\prime c^\prime
d^\prime}$ that is simply totally symmetric but with arbitrary index
values. The constraint can then be imposed by contracting with
$P^{abcd}_{a^\prime b^\prime c^\prime d^\prime }
=\delta_{aa^\prime}^{bcd}\delta_{bb^\prime}^{cd}\delta_{cc^\prime}^{d}\delta_{dd^\prime}$.
Note that $\delta^{cd}_{bb^\prime} = \delta_{bb^\prime}-\delta_{bb^\prime
c} - \delta_{bb^\prime d } + \delta_{bb^\prime dc}$. But the utility of
this formalism is now clear: if one wishes to evaluate a relation such as
$\delta_{bb^\prime}^{cd}\delta_{cc^\prime}^{d}\delta_{dd^\prime}$, the last
term in $\delta^{cd}_{bb^\prime}$, namely $\delta_{bb^\prime dc}$, can be
dropped since it gives zero when contracted with
$\delta_{cc^\prime}^{d}\delta_{dd^\prime}=(\delta_{cc^\prime}\delta_{dd^\prime}-\delta_{cc^\prime
dd^\prime})$. Thus, in conclusion, as long as we calculate expressions like
$\delta_{aa^\prime}^{bcd}\delta_{bb^\prime}^{cd}\delta_{cc^\prime}^{d}\delta_{dd^\prime}$
from right to left, it is sufficient to take $\delta_{b b^\prime}^{a_1 a_2
\ldots a_n} =\delta_{b b^\prime}-\sum_{i = 0 }^n \delta_{bb^\prime a_i}$. This allows us to algorithmically impose constraints on tensors.

\section{\texorpdfstring{$\boldsymbol{\langle ZZZZ \rangle}$}{\textlangle
  ZZZZ\textrangle} sum rules for \texorpdfstring{$\boldsymbol{n \geq
  4}$}{n>= 4}}
Due to the rather extended size of the sum rules derived from the crossing
equation corresponding to $\langle ZZZZ \rangle$ for $n \geq 4$ we attach
them in an auxiliary file.

\newpage

\end{appendices}

\bibliography{MN_mixed_correlator}

\end{document}